\documentclass[twocolumn
]{jpsj2}
\usepackage[dvips]{color}

\title{
Charge Order with Structural Distortion in Organic Conductors :
Comparison between $\theta$-(ET)$_2$RbZn(SCN)$_4$ and
$\alpha$-(ET)$_2$I$_3$
}

\author{
Yasuhiro \textsc{Tanaka$^1$}\thanks{yasuhiro@ims.ac.jp}
and Kenji \textsc{Yonemitsu$^{1,2}$}\thanks{kxy@ims.ac.jp}
}

\inst{
$^1$Institute for Molecular Science, Okazaki, Aichi 444-8585\\
$^2$Graduate University for Advanced Studies, Okazaki, Aichi\ 444-8585
}

\date{\today}

\abst{
Charge ordering with structural distortion in quasi-two-dimensional
organic conductors $\theta$-(ET)$_2$RbZn(SCN)$_4$ (ET=BEDT-TTF) and
$\alpha$-(ET)$_2$I$_3$ 
is investigated theoretically. By using the Hartree-Fock
approximation for an extended Hubbard model which includes both on-site
and intersite Coulomb interactions together with Peierls-type
electron-lattice couplings, we examine the role of lattice degrees of 
freedom on charge order. 
It is found that the experimentally observed, horizontal charge order is
stabilized by lattice distortion in both compounds.
In particular, the lattice effect is crucial to the realization of the
charge order in $\theta$-(ET)$_2$RbZn(SCN)$_4$, while the peculiar band
structure whose symmetry is lower than that of
$\theta$-(ET)$_2$RbZn(SCN)$_4$ in the metallic phase is also an
important factor in $\alpha$-(ET)$_2$I$_3$ together with the lattice
distortion. 
For $\alpha$-(ET)$_2$I$_3$, we obtain a phase transition
from a charge-disproportionated metallic phase to the horizontal charge
order with lattice modulations, which is consistent
with the latest X-ray experimental result. 
}

\kword{charge order, organic conductor, extended Hubbard model,
electron-lattice coupling, Hartree-Fock approximation}

\begin{document}
\sloppy
\maketitle

\section{Introduction}
The family of quasi-two dimensional organic conductors (ET)$_2$X
(ET=BEDT-TTF) exhibit a variety of interesting physical properties
such as superconductivity, magnetism and charge
order (CO)\cite{Ishiguro,Seo_Rev}. (ET)$_2$X
consist of stacking layers of donor ET molecules and monovalent X
anions. CO phenomena are observed in several members of (ET)$_2$X, 
which have a 3/4-filled $\pi$-electron band, and have been studied
intensively since they have an important role to our understanding of 
electron correlation effects in low-dimensional systems.

$\theta$-(ET)$_2$RbZn(SCN)$_4$\cite{Miyagawa,Chiba} and
$\alpha$-(ET)$_2$I$_3$\cite{Takano1,Takano2} are 
typical compounds that are known to exhibit CO.
$\theta$-(ET)$_2$RbZn(SCN)$_4$ shows a
first-order metal-insulator transition at $T_c=200$K and a spin gap
behavior at low temperatures much below $T_c$\cite{Mori1}. The
transition is accompanied by lattice distortion which changes its
structure from
the $\theta$-type in the metallic phase to the so-called $\theta_d$-type in
the insulating phase. The CO formation below $T_c$ has been
directly observed by NMR experiments\cite{Miyagawa,Chiba}. Several
experiments\cite{Tajima,Wang,Yamamoto,Watanabe1,Watanabe2} such as Raman
scattering\cite{Yamamoto} and X-ray scattering\cite{Watanabe1,Watanabe2} 
measurements show that the horizontal-type CO is formed in this
compound. In the metallic phase, X-ray experiments 
indicate a short-range CO with long
periodicity, which is different from the horizontal-stripe state at low
temperatures\cite{Watanabe1,Watanabe2}. 
The NMR measurement also shows slow fluctuations of CO in the metallic
state\cite{Chiba2}.
Similar charge fluctuations are
observed in $\theta$-(ET)$_2$CsZn(SCN)$_4$, which suggests the
coexistence of charge modulations with different wave vectors\cite{Watanabe4,Nogami}
although no long-range CO has been observed in this compound.

On the other hand, $\alpha$-(ET)$_2$I$_3$ shows a metal-insulator
transition at $T=135$K\cite{Bender}, and a spin
gap behavior is observed below $T_c$ by the magnetic susceptibility
measurement\cite{Rothamael}. Charge ordering in $\alpha$-(ET)$_2$I$_3$
has been suggested theoretically by the Hartree-Fock
approximation\cite{Kino1,Kino2} for the model that takes account of the
full anisotropy of transfer integrals in this compound. Experimentally,
the CO among ET molecules is confirmed by NMR
experiments\cite{Takano1,Takano2} and Raman
spectroscopy\cite{Woj}. Moreover, it has been
observed that the charge disproportionation exists even in the metallic
state\cite{Woj,Moroto}. The horizontal-type CO has been recently and
directly observed by the X-ray scattering study\cite{Kakiuchi}, which
also shows that the transition is accompanied by a structural
distortion\cite{Emge,Endres,Nogami2}.

Although the compounds $\theta$-(ET)$_2$RbZn(SCN)$_4$ and
$\alpha$-(ET)$_2$I$_3$ have COs with very similar spatial patterns at
low temperatures, the natures of the CO transitions are quite
different. In
$\theta$-(ET)$_2$RbZn(SCN)$_4$, the transition is of first order
and accompanied with a large lattice distortion and large
discontinuity. 
On the other hand, the
lattice distortion in $\alpha$-(ET)$_2$I$_3$ is relatively small and,
although
the transition is of first order, the hysteresis of the specific heat
is substantially smaller than that of
$\theta$-(ET)$_2$RbZn(SCN)$_4$\cite{Nishio,Fortune}. In fact, the recently
observed photoinduced melting of CO in these compounds\cite{Iwai}
shows a clear difference in the photoinduced dynamics, which is considered to
originate from different roles of electron-lattice couplings in COs.

Theoretically, CO phenomena have been investigated by using the 
extended Hubbard model including on-site ($U$) and intersite ($V_{ij}$)
Coulomb
interactions\cite{Seo,Mckenzie,Mori2,Kaneko,Clay,Merino,Watanabe3,Kuroki,Hotta1,Hotta2
,Udagawa,Tanaka,Miyashita}.
The stability of various CO patterns in (ET)$_2$X has been discussed
first within the Hartree approximation by considering the realistic band
structures of (ET)$_2$X\cite{Seo_Rev,Seo}. The horizontal-stripe CO is
shown to be stabilized in $\theta$-(ET)$_2$RbZn(SCN)$_4$ and also in
$\alpha$-(ET)$_2$I$_3$. In particular, the horizontal CO is shown to be
more stable in the $\theta_d$-type structure than in the $\theta$-type one,
which suggests the lattice effects have an important role to realize the
horizontal CO in $\theta$-(ET)$_2$RbZn(SCN)$_4$. For $\theta$-(ET)$_2$X,
COs with long periodicity are also considered by several
authors\cite{Mori2,Kaneko,Watanabe3,Kuroki,Hotta1,Hotta2,Udagawa,Tanaka} and
they discussed their relevance to anomalous charge fluctuations in the
metallic phase. 

In the previous paper\cite{Tanaka}, we have studied the effects of 
lattice distortion on CO in $\theta$-(ET)$_2$X within the Hartree-Fock
approximation by taking account of three Peierls-type
electron-lattice couplings explicitly, which are
deduced from the X-ray scattering experiment\cite{Watanabe5}. The
results show that the horizontal CO is stabilized by the lattice
distortion, which results in the structural change from the $\theta$-type
to the $\theta_d$-type. This is consistent with the experiments on
$\theta$-(ET)$_2$RbZn(SCN)$_4$ and also consistent with the recent
numerical exact-diagonalization study\cite{Miyashita}.

In this paper, we discuss the role of each electron-lattice coupling to
the formation of CO in detail for
$\theta$-(ET)$_2$RbZn(SCN)$_4$. Furthermore, we
consider the lattice effects in $\alpha$-(ET)$_2$I$_3$ and compare the
results with those for $\theta$-(ET)$_2$RbZn(SCN)$_4$. It is found that 
although the lattice degrees of freedom are crucial to the realization of 
horizontal CO in $\theta$-(ET)$_2$RbZn(SCN)$_4$, the effects are
relatively small in $\alpha$-(ET)$_2$I$_3$ where the band structure and
Coulomb interactions have an important role. At finite temperatures, we
compare the free energies of various CO patterns and 
investigate the CO phase transitions in both compounds. 
This paper is organized as follows. In \S 2, the extended Hubbard model
and the Peierls-type electron-lattice couplings for
$\theta$-(ET)$_2$RbZn(SCN)$_4$ and $\alpha$-(ET)$_2$I$_3$ are
introduced. We show the results of the Hartree-Fock calculations at
zero and finite temperatures and discuss their relevance to the
experimental results in \S 3. \S 4 is devoted to the summary and
conclusion.

\section{Model and Method}
\subsection{Formulation}
We consider the extended Hubbard model written as, 
\begin{equation}
\begin{split}
{\it H}=&\sum_{\langle ij \rangle\sigma}(t_{i,j}+ \alpha_{i,j}u_{i,j})
(c^{\dagger}_{i\sigma}c_{j\sigma}+\rm{h.c})\\
&+U\sum_{i}n_{i\uparrow}n_{i\downarrow}+\sum_{\langle ij \rangle}
 V_{i,j}n_{i}n_{j}+\sum_{\langle ij \rangle}\frac{K_{i,j}}{2}u^{2}_{i,j}\ ,
\end{split}
\end{equation}
where $\langle ij\rangle$ represents the summation over the pairs of 
neighboring sites, $c^{\dagger}_{i\sigma}(c_{i\sigma})$ denotes the
creation (annihilation) operator for an electron with spin $\sigma$ at
the $i$th site, $n_{i\sigma}=c^{\dagger}_{i\sigma}c_{i\sigma}$, and
$n_{i}=n_{i\uparrow}+n_{i\downarrow}$. The electron density is fixed at 
3/4 filling. The electron-lattice coupling
constant, lattice displacement and elastic constant are denoted by
$\alpha_{i,j}$, $u_{i,j}$ and $K_{i,j}$, respectively.
For the intersite Coulomb interactions $V_{i,j}$, we consider nearest
neighbor interactions $V_c$ for the vertical direction and $V_p$ for
the diagonal direction as shown in Fig. 1(a). 
We further rewrite the parameters for the lattice degrees of freedom by 
introducing new variables as 
$y_{i,j}=\alpha_{i,j}u_{i,j}$ and $s_{i,j}=\alpha_{i,j}^{2}/K_{i,j}$.
The actual form of $y_{i,j}$ together with the transfer integrals 
$t_{i,j}$ for $\theta$-(ET)$_2$RbZn(SCN)$_4$ and $\alpha$-(ET)$_2$I$_3$
are given below. We apply the Hartree-Fock approximation
\begin{equation}
\begin{split}
n_{i\sigma}n_{j\sigma^{\prime}}\rightarrow
&\langle n_{i\sigma}\rangle n_{j\sigma^{\prime}}+
n_{i\sigma}\langle n_{j\sigma^{\prime}}\rangle -\langle
 n_{i\sigma}\rangle \langle n_{j\sigma^{\prime}}\rangle \\
&-\langle c_{i\sigma}^{\dagger}c_{j\sigma^{\prime}}\rangle
c_{j\sigma^{\prime}}^{\dagger}c_{i\sigma}-c_{i\sigma}^{\dagger}c_{j\sigma^{\prime}}
\langle c_{j\sigma^{\prime}}^{\dagger}c_{i\sigma}\rangle \\
&+\langle c_{i\sigma}^{\dagger}c_{j\sigma^{\prime}}\rangle \langle
 c_{j\sigma^{\prime}}^{\dagger}c_{i\sigma}\rangle
\end{split}
\end{equation}
to eq. (1) and the obtained Hamiltonian is diagonalized in $k$-space by
assuming unit cells of various mean-field order parameters. We consider
four types of order parameters according to the alignment of hole-rich
molecules, namely, 3-fold,
diagonal, horizontal and vertical COs which are
schematically shown in Fig. 2. As for the spin degrees of freedom, we
use three spin configurations in each stripe-type CO which are
identical to those of ref. 29, while the antiferromagnetic state between
hole-rich and -poor sites is considered in the 3-fold CO. The
ground-state energy is calculated by solving the mean-field equation
self-consistently together with the lattice displacements,
which are determined by the Hellmann-Feynman theorem.
The energy per site is given by
\begin{equation}
\hspace*{-1.0cm} 
\begin{split}
E &=
 \frac{1}{N}\Bigl(\sum_{l\bf{k}\sigma}E_{l\bf{k}\sigma}n_{F}(E_{l\bf{k}\sigma})-
U\sum_{i}\langle n_{i\uparrow}\rangle \langle n_{i\downarrow}\rangle \\
&-\sum_{\langle ij\rangle}V_{ij}\langle n_{i}\rangle \langle
 n_{j}\rangle +\sum_{\langle ij\rangle \sigma}V_{ij}\langle
 c_{i\sigma}^{\dagger}c_{j\sigma}\rangle \langle
 c_{j\sigma}^{\dagger}c_{i\sigma}\rangle \\
&+\sum_{\langle ij\rangle}\frac{y_{ij}^{2}}{2s_{ij}}\Bigr)\ ,
\end{split}
\end{equation}
where {\it l}, $E_{l\bf{k}\sigma}$ and $n_{F}$ is a band index, an
energy eigenvalue of the mean-field Hamiltonian and the Fermi
distribution function, respectively. $N$ is the total number of
sites. We notice that the presence or absence of the Fock terms in
eq. (2) does not alter the qualitative results including the lattice
displacements although the energy of each self-consistent solution is
lowered by the Fock terms.
\begin{figure}[h]
\begin{center}
\includegraphics[width=9.0cm]{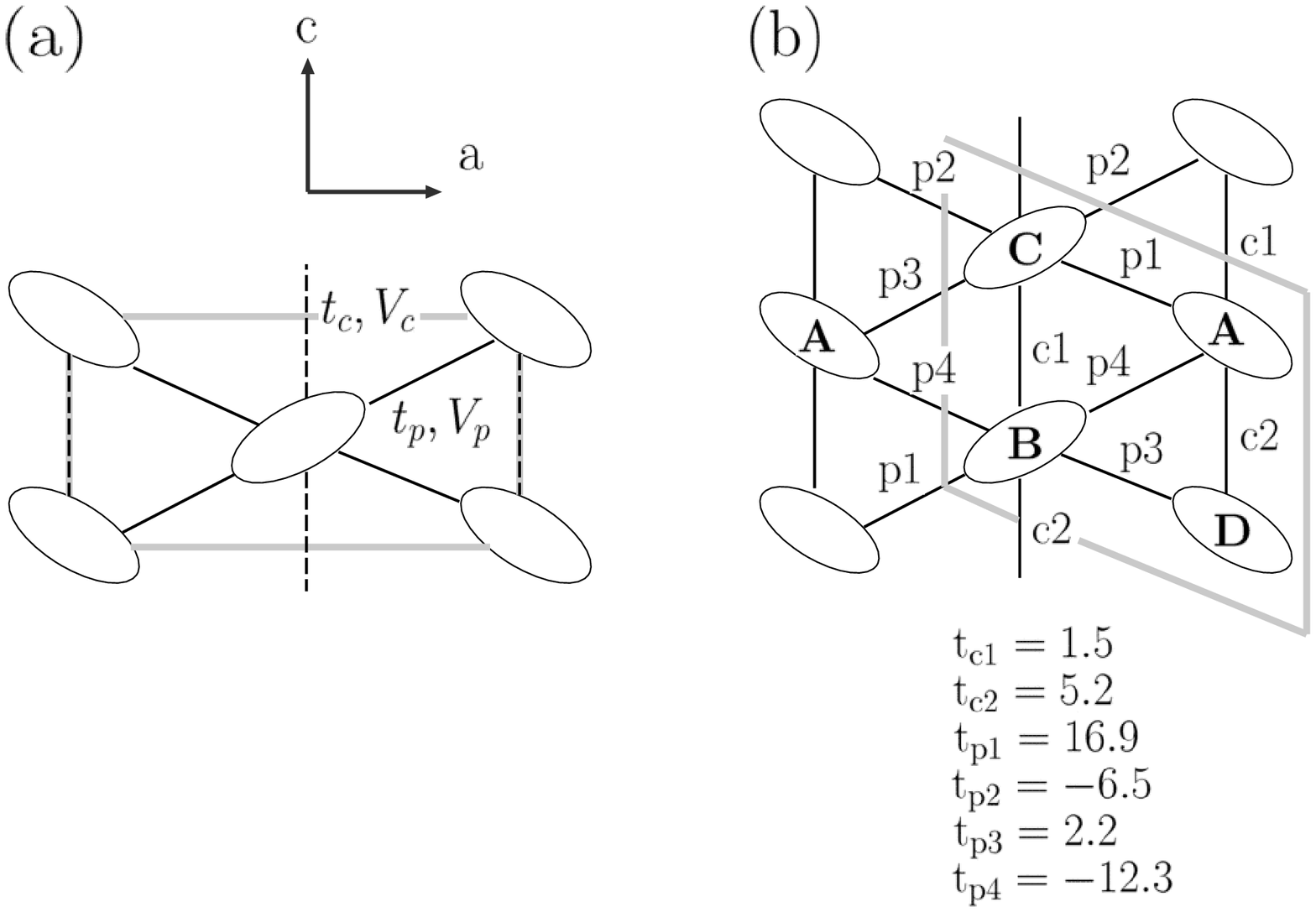}
 \includegraphics[width=9.0cm]{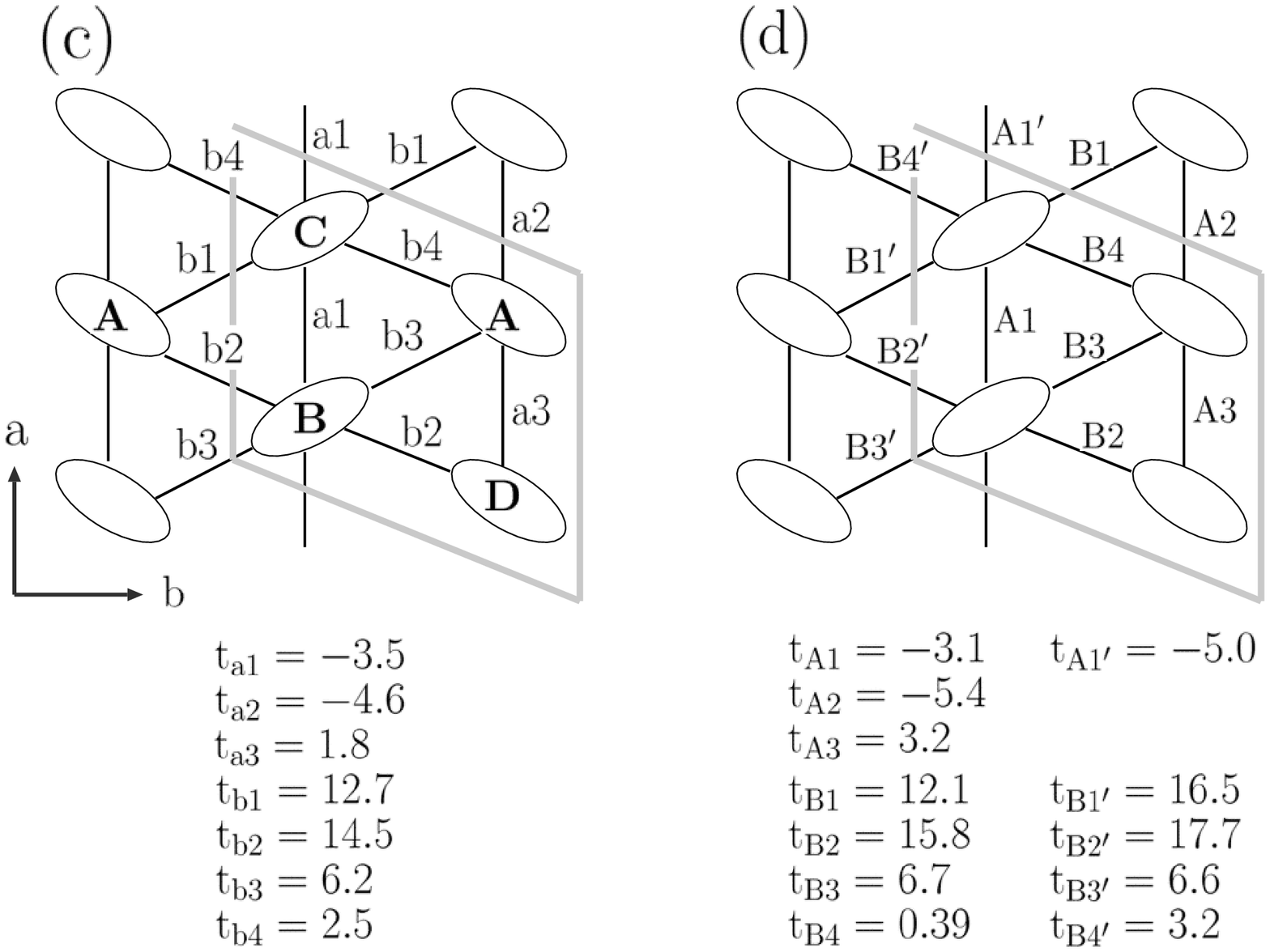}
\caption{Schematic representations of the structures of
 (a)$\theta$-(ET)$_2$X, (b)$\theta$-(ET)$_2$RbZn(SCN)$_4$ in the CO
 phase, (c)$\alpha$-(ET)$_2$I$_3$ in the metallic phase, and (d)
$\alpha$-(ET)$_2$I$_3$ in the CO phase. The gray solid lines indicate
 the unit cell. For (b), (c), and (d), the transfer integrals
 estimated by the extended H$\ddot{\rm u}$ckel
 method\cite{Watanabe1,Kakiuchi} are also shown. For
 $\alpha$-(ET)$_2$I$_3$, the site indices A, B, C, and D in our notation
 correspond to A, B, C, and ${\rm A}^{\prime}$, respectively, in the
 conventional notation\cite{Mori4}.}
\end{center}
\end{figure}
\begin{figure}[h]
\begin{center}
\includegraphics[width=8.0cm]{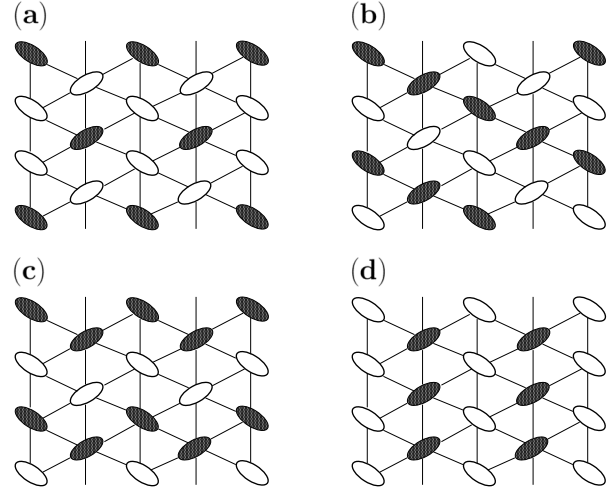}
\caption{Spatial patterns of COs considered in the Hartree-Fock
 calculations. (a)3-fold, (b)diagonal-stripe, (c)horizontal-stripe, and
(d)vertical-stripe COs. The hole-rich and hole-poor sites are denoted by
the solid and open ellipses, respectively.}
\end{center}
\end{figure}

At finite temperatures we calculate the free energy within the Hartree-Fock
approximation. The free energy per site is written as
\begin{equation}
\begin{split}
F &=
 \frac{1}{N}\Bigl(\mu N_{tot}-\frac{1}{\beta}\sum_{l\bf{k}\sigma}\ln
 (1+\exp \{-\beta (E_{l\bf{k}\sigma}-\mu)\}) \\
&-U\sum_{i}\langle n_{i\uparrow}\rangle \langle n_{i\downarrow}\rangle 
-\sum_{\langle ij\rangle}V_{ij}\langle n_{i}\rangle \langle n_{j}\rangle
 \\
&+\sum_{\langle ij\rangle \sigma}V_{ij}\langle
 c_{i\sigma}^{\dagger}c_{j\sigma}\rangle \langle
 c_{j\sigma}^{\dagger}c_{i\sigma}\rangle
 +\sum_{\langle ij\rangle}\frac{y_{ij}^{2}}{2s_{ij}}\Bigr)\ ,
\end{split}
\end{equation}
where $\mu$, $N_{tot}$ and $\beta$ is the chemical potential, total
number of electrons and inverse of temperature, respectively. 

\subsection{$\theta$-(ET)$_2$RbZn(SCN)$_4$}
The structure of $\theta$-(ET)$_2$RbZn(SCN)$_4$ in the metallic phase
and that in the CO phase are shown in Figs. 1(a) and 1(b),
respectively. In the metallic
phase, there are two kinds of transfer integrals $t_p$ and $t_c$ for 
diagonal and vertical bonds, while the CO transition doubles the unit
cell in the $c$-direction and six transfer integrals exist at low
temperatures. In order to take account of the lattice effects, we
consider three kinds of electron-lattice
couplings, $s_{i,j}$: $s_c$, $s_a$, and $s_{\phi}$
that originate from the $c$- and $a$-axis molecular translations and
molecular rotation, respectively. They have been introduced in the previous
papers\cite{Tanaka,Miyashita}, being suggested by the X-ray experiment
that shows rather complicated
displacements of ET molecules through the CO transition\cite{Watanabe1}. 
First, the $c$-axis translation alternates $t_c$. This modulation gives
$t_{c1}$ and $t_{c2}$ in Fig. 1(b). Similarly, the $a$-axis translation
induces the
modulation of $t_{p1}$ and $t_{p3}$. Finally, the rotational degrees of
freedom are taken into account by the changes in $t_{p2}$ and
$t_{p4}$. This type
of modulation is expected to be important since the horizontal CO is
formed on the $t_{p4}$ chains with hole-rich molecules. In fact, the 
experimental estimation\cite{Watanabe1} indicates that the dependence of
transfer integrals on 
a relative angle (called elevation angle\cite{Watanabe1}) of neighboring
ET molecules is large and allows uniformly decreasing
(increasing) $|t_{p2}|$ ($|t_{p4}|$) on the horizontally connected bonds.
These three kinds of electron-lattice couplings cause the modulations
$y_{i,j}$ of transfer integrals that are experimentally observed. For
simplicity, they are assumed to be independent and we do not consider
any other electron-lattice coupling.
Then, the transfer integrals in the distorted structure are given by, 
\begin{equation}
\begin{split}
&t_{c1}=t_{c}+y_{c}\ , \\
&t_{c2}=t_{c}-y_{c}\ ,\\
&t_{p1}=t_{p}+y_{a}\ , \\
&t_{p2}=t_{p}-y_{\phi}\ ,\\ 
&t_{p3}=t_{p}-y_{a}\ ,\\
&t_{p4}=t_{p}+y_{\phi}\ ,
\end{split}
\end{equation}
where $y_{c}$, $y_{a}$, and $y_{\phi}$ are modulations due to $s_{c}$,
$s_{a}$, and $s_{\phi}$, respectively. Here the signs in eq. (5) are
chosen so that the transfer integrals
realized experimentally at low temperatures correspond to $y_{l}>0$ for
$l=c,\ a$ and $\phi$\cite{Tanaka}. In the calculations,
we use $t_{p}=0.1 {\rm (eV)}$ and $t_{c}=-0.04$ in the high temperature
phase
for $t_{i,j}$ in eq. (1).
Then, the modulations
$y_{c}$, $y_{a}$, and $y_{\phi}$ are self-consistently determined for
each set of electron-lattice coupling constants.

\subsection{$\alpha$-(ET)$_2$I$_3$}
The transfer integrals for $\alpha$-(ET)$_2$I$_3$ in the high- and
low-temperature phases are shown in Figs. 1(c) and 1(d), respectively. 
The unit cell contains four molecules in both phases.
According to the X-ray structural analysis\cite{Kakiuchi},
sites A and D are equivalent to each other owing to the inversion
symmetry in the metallic phase, while the symmetry breaks 
 below the CO transition temperature.
Sites A and B (C and D) become hole-rich (hole-poor) in the horizontal
CO state.
The large transfer integrals, $t_{b1}$ and $t_{b2}$, form a
one-dimensional zigzag chain in the conducting layer. The extended
H$\ddot{\rm u}$ckel calculation indicates that they gradually increase with
decreasing temperature\cite{Kakiuchi}. At the CO transition, $t_{b1}$
changes into $t_{\rm B1}$ and $t_{\rm B1^{\prime}}$ in Fig. 1(d), while 
$t_{b2}$ becomes $t_{\rm B2}$ and $t_{\rm B2^{\prime}}$. The other
transfer integrals with smaller values also change at the
transition. The lattice modulation in this system is suggested to come
mainly from the molecular rotation rather than the translational shifts
of the molecules\cite{Kakiuchi}.
\begin{figure}
\begin{center}
\includegraphics[width=8.0cm,clip]{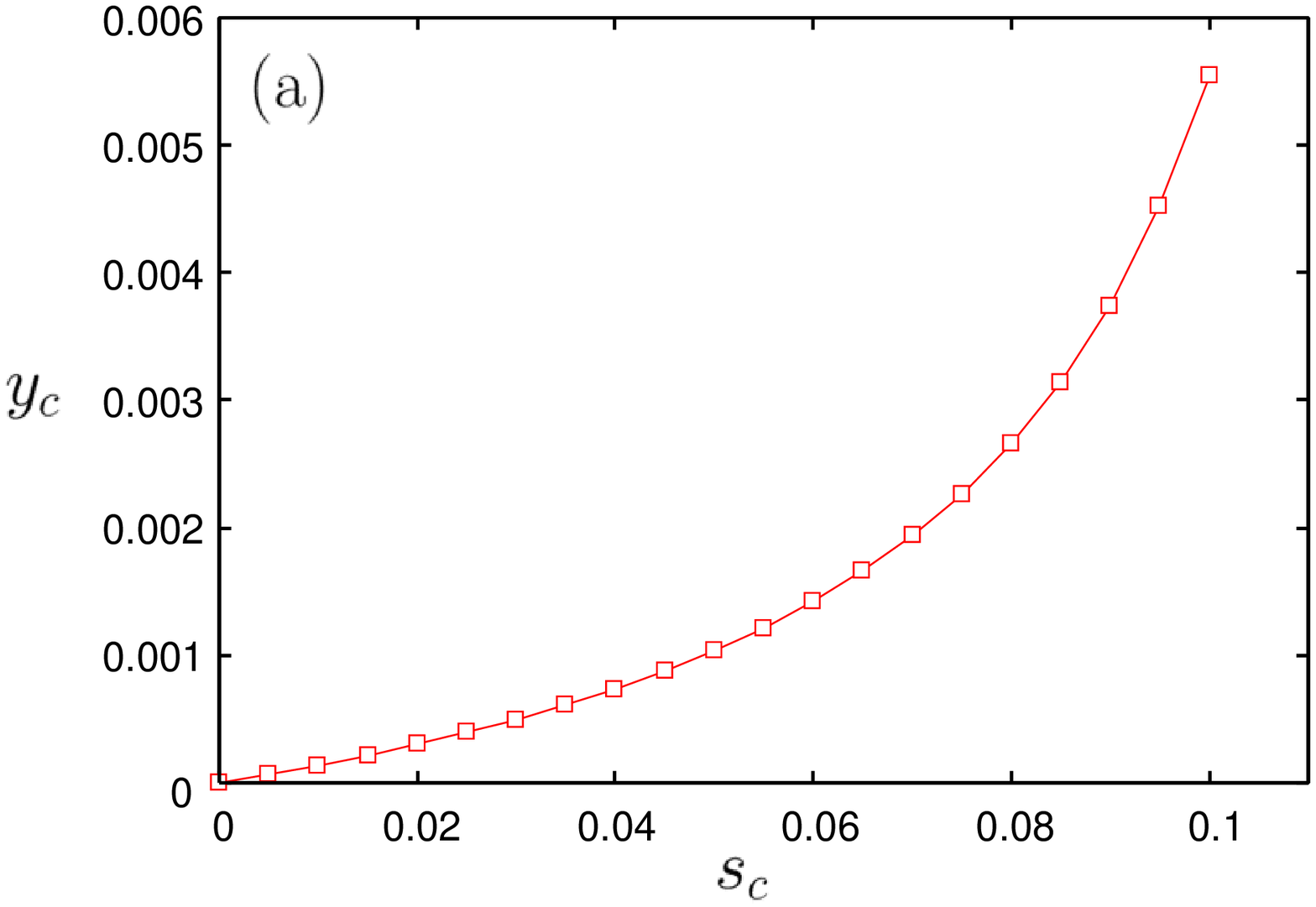}\\
\vspace{0.3cm}
\includegraphics[width=8.0cm,clip]{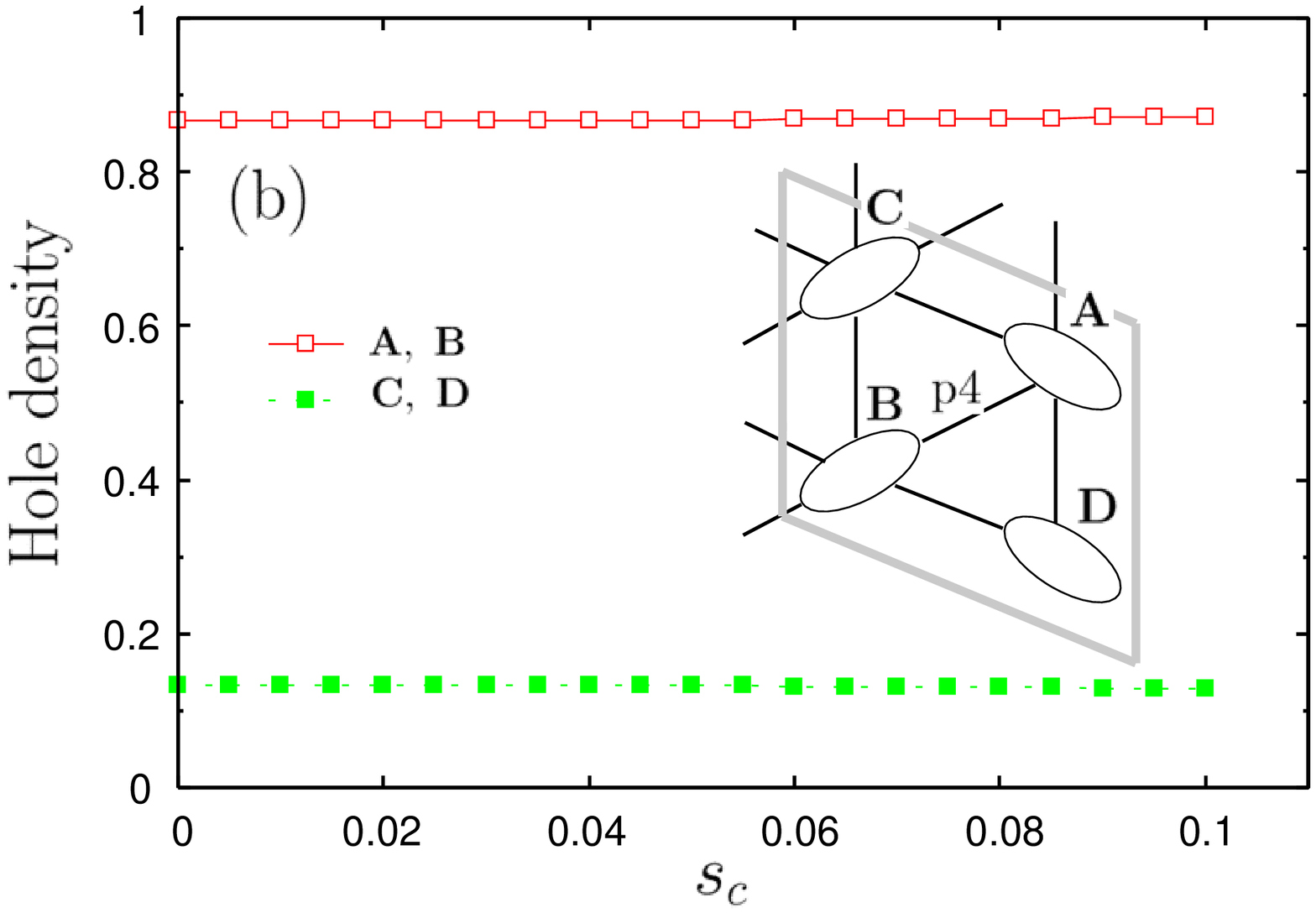}
\end{center}
\caption{(Color online) (a)Modulation $y_{c}$ and (b)hole densities as a
 function of $s_{c}$ for $U=0.7$, $V_c/U=0.35$, and $V_p/V_c=0.7$.}
\end{figure}
\begin{figure}
\begin{center}
\includegraphics[width=8.0cm,clip]{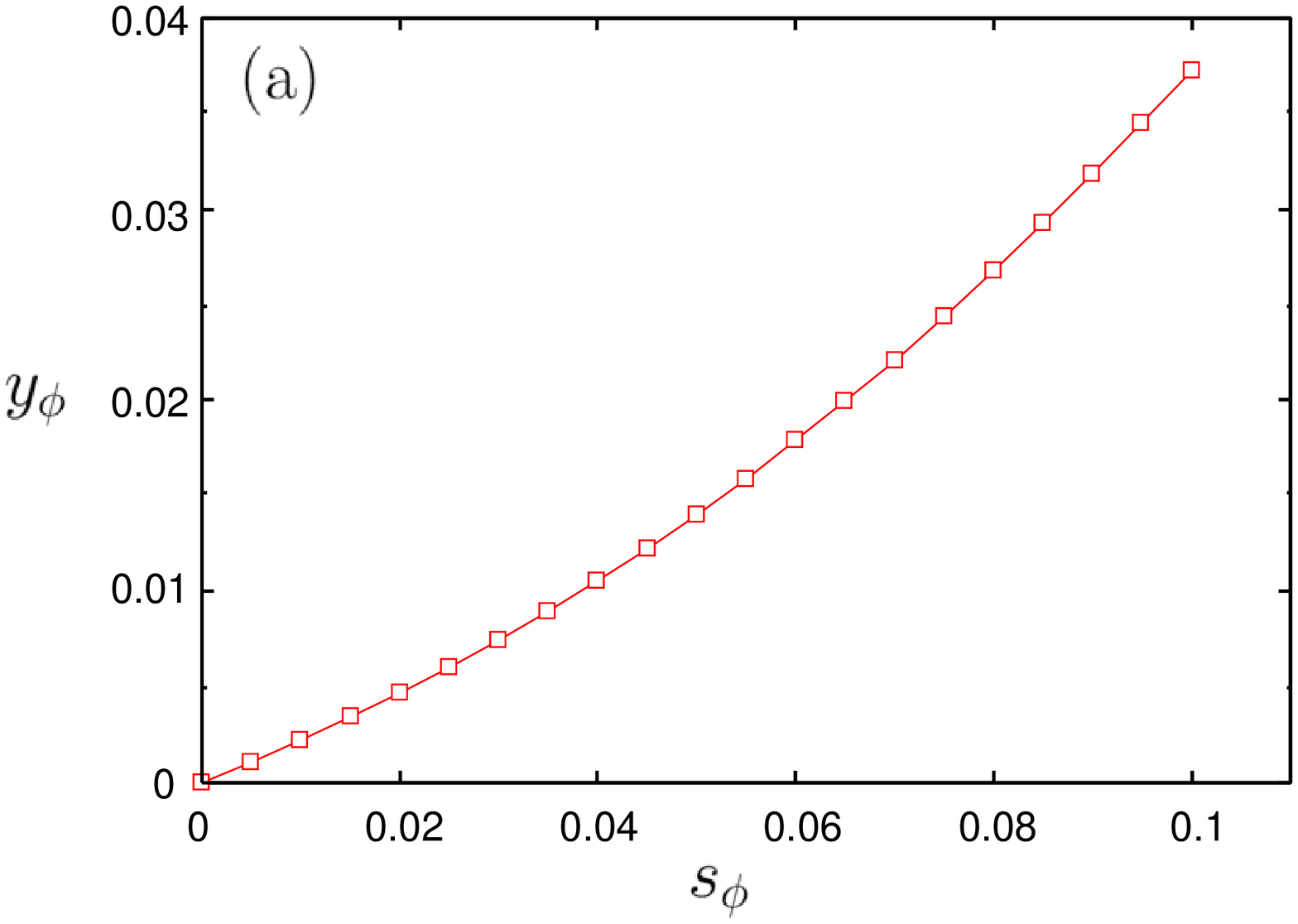}\\
\vspace{0.3cm}
\includegraphics[width=8.0cm,clip]{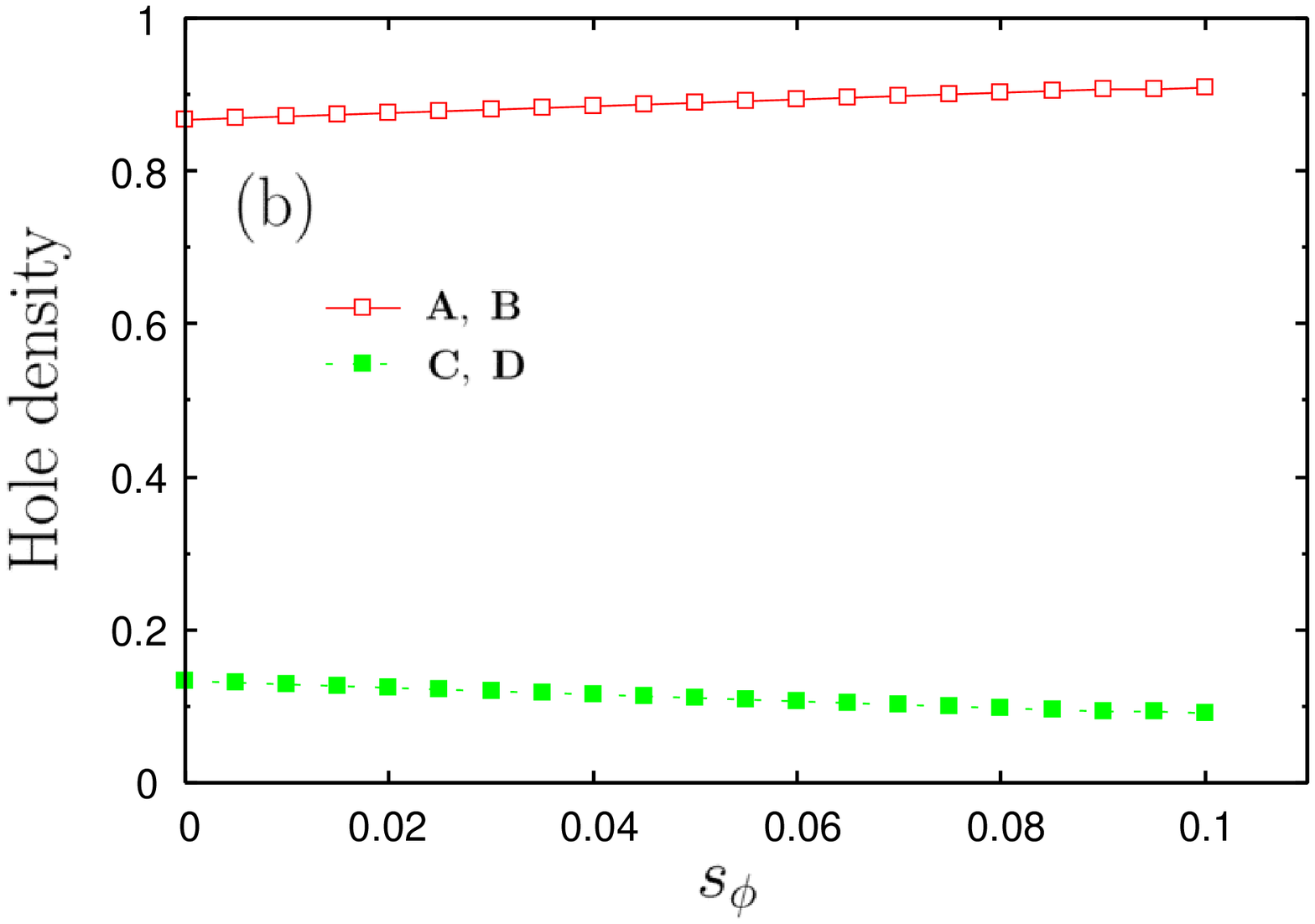}
\end{center}
\caption{(Color online) (a)Modulation $y_{\phi}$ and (b)hole densities
 as a function of $s_{\phi}$ for $U=0.7$, $V_c/U=0.35$, and
 $V_p/V_c=0.7$.}
\end{figure}

In the present paper, we investigate the lattice effects on the system
by taking account of the modulations in $t_{b1}$ and 
$t_{b2}$. We do not consider the modulations in smaller transfer
integrals than
the above ones. These effects will be mentioned in the following
section. As in the case of $\theta$-(ET)$_2$RbZn(SCN)$_4$, we let 
$s_{b1}$ and $s_{b2}$ denote the electron-lattice couplings that change
$t_{b1}$ and $t_{b2}$, respectively. The corresponding
modulations are written as $y_{b1}$ and $y_{b2}$. The transfer integrals
in the low temperature phase are then written as
\begin{equation}
\begin{split}
&t_{{\rm B1^{\prime}}}= t_{b1}+y_{b1}\ , \\
&t_{{\rm B1}}=t_{b1}-y_{b1}\ ,\\
&t_{{\rm B2^{\prime}}}=t_{b2}+y_{b2}\ , \\
&t_{{\rm B2}}=t_{b2}-y_{b2}\ ,
\end{split}
\end{equation}
where the signs in eq. (6) are again chosen so as for positive values to
correspond to the modulations realized in $\alpha$-(ET)$_2$I$_3$. 
The other transfer integrals are assumed to be unchanged.
For the values of $t_{i,j}$ in eq. (1) with which the Hartree-Fock
calculations are carried out, we use those in the high temperature phase
that are shown in Fig. 1(c).
 
\section{Results and Discussions}
\subsection{$\theta$-(ET)$_2$RbZn(SCN)$_4$}
First, let us consider the effect of each electron-lattice coupling on
the horizontal CO that is experimentally observed in
$\theta$-(ET)$_2$RbZn(SCN)$_4$. Figures 3, 4 and 5 show the lattice
displacements and order parameters as a function of the
electron-lattice coupling when we assume the horizontal CO
as the mean-field order parameter for $U=0.7$, $V_c/U=0.35$, and
$V_{p}/V_{c}=0.7$. Note that the obtained CO state is not the ground
state for small electron-lattice couplings, as we will discuss
later. From Figs. 3(a) and 4(a), we can see that $y_{c}$ and $y_{\phi}$
increase
with increasing $s_{c}$ and $s_{\phi}$, respectively. The modulations
grow linearly for small $s_{c}$ and $s_{\phi}$, which indicates that
the horizontal CO is stabilized by these lattice modulations.
The hole density at each site in 
the unit cell shows
that the disproportionation is enlarged by $s_{\phi}$ although the
dependence is weak in the case of $s_{c}$. On the other hand, the
behavior of $y_{a}$
is different from the above two modulations, as shown in Fig. 5. $y_{a}$
becomes nonzero when $s_{a}$ exceeds some value. 
With the increase of $y_{a}$, the order parameter tends to decrease,
which shows the $y_{a}$ modulation suppresses the charge
disproportionation in the horizontal CO. The energy gain by these
electron-lattice couplings is
shown in Fig. 6. We find that the effect of $s_{\phi}$ is the
largest. Since the horizontal stripe CO is formed on the hole-rich sites
that are connected by the $t_{p4}$ chain, the $y_{\phi}$ modulation
lowers the energy through the exchange coupling between neighboring
spins on the stripe. The energy gain due to spin fluctuations is also
obtained by the perturbation theory from the strong-coupling limit,
i.e., $t_{i,j}=0$\cite{Miyashita}.
\begin{figure}
\begin{center}
\includegraphics[width=8.0cm,clip]{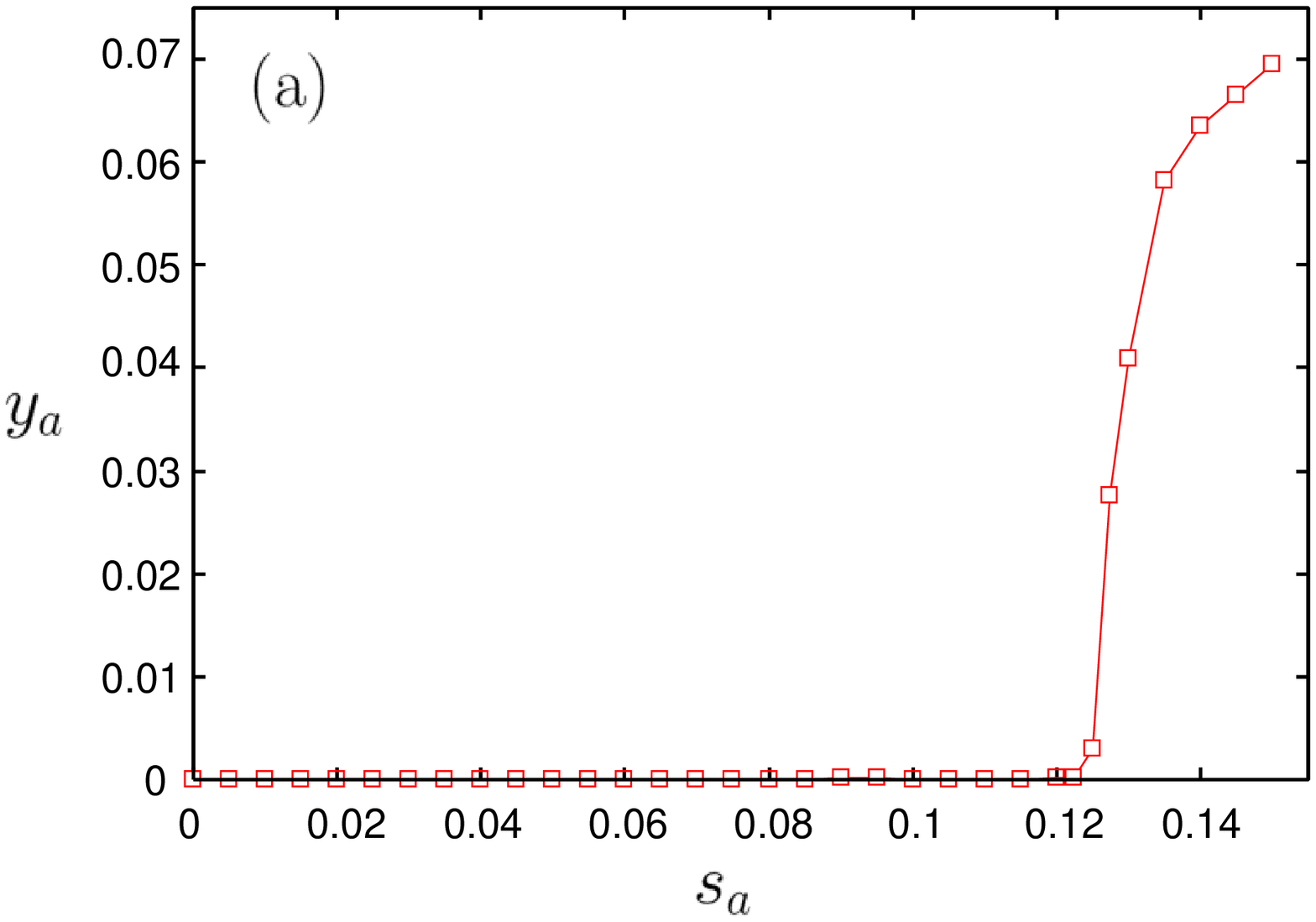}\\
\vspace{0.3cm}
\includegraphics[width=8.0cm,clip]{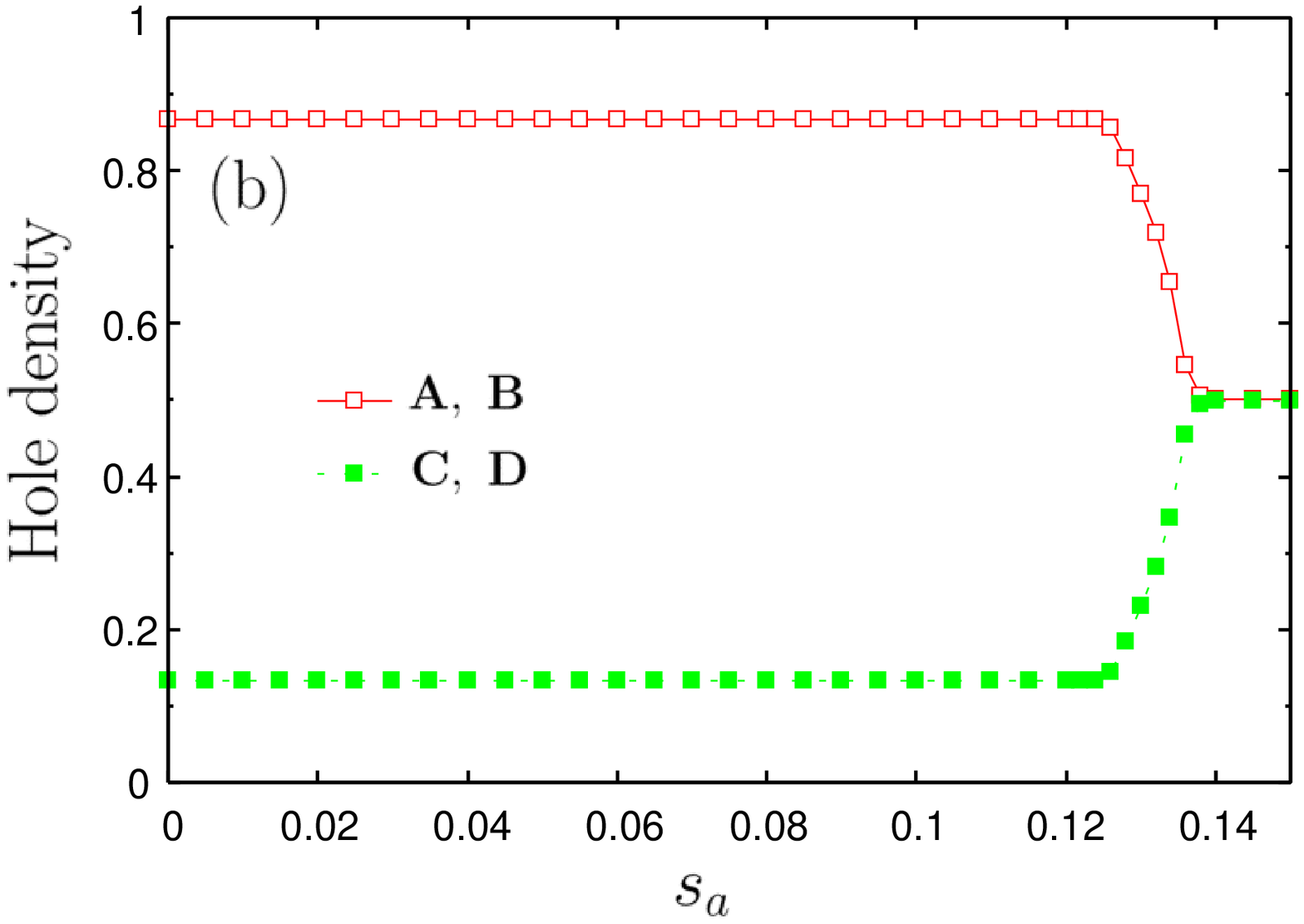}
\end{center}
\caption{(Color online) (a)Modulation $y_{a}$ and (b)hole densities as a
 function of $s_{a}$ for $U=0.7$, $V_c/U=0.35$, and $V_p/V_c=0.7$.}
\end{figure}
\begin{figure}
\begin{center}
\includegraphics[width=8.0cm,clip]{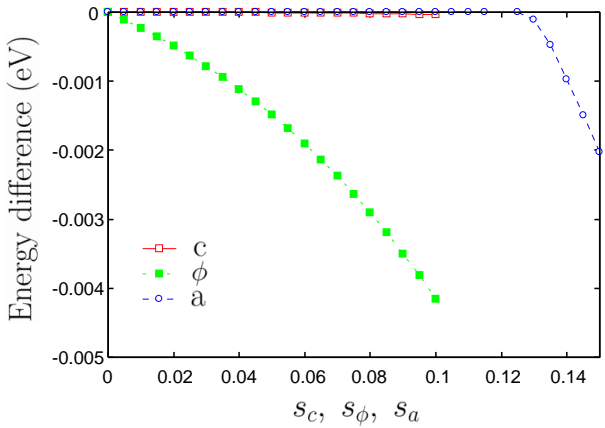}
\end{center}
\caption{(Color online) Energy gain due to each electron-lattice
 coupling in the horizontal CO for $U=0.7$, $V_c/U=0.35$, and
 $V_p/V_c=0.7$.}
\end{figure}

Next, we discuss the stability of various CO patterns. The ground-state
energies as a function of $V_p/V_c$ are compared in Fig. 7(a). In this
figure, the energy of the 3-fold CO without electron-lattice couplings
is set at zero.
We have shown only the
lowest energy state for each CO pattern among different spin
configurations. In the absence of electron-lattice couplings, the 3-fold
CO with a ferrimagnetic spin configuration is the most
favorable in the nearly isotropic region, $V_p/V_c\sim 1$, while
the diagonal CO whose spin configuration is antiferromagnetic along the
stripe and between stripes on the $c$-axis is stable when $V_p/V_c$ is
small, $V_p/V_c<0.7$\cite{Kaneko,Tanaka}. For the horizontal CO,
we plotted the energy
of the state that is antiferromagnetic along the stripe and
ferromagnetic between stripes on the $c$-axis. The state that is
antiferromagnetic on the $c$-axis has a close energy and is nearly
degenerate with the above state. Without electron-lattice coupling, the
energy of the horizontal CO cannot be the lowest in any region.
\begin{figure}
\begin{center}
\includegraphics[width=8.0cm,clip]{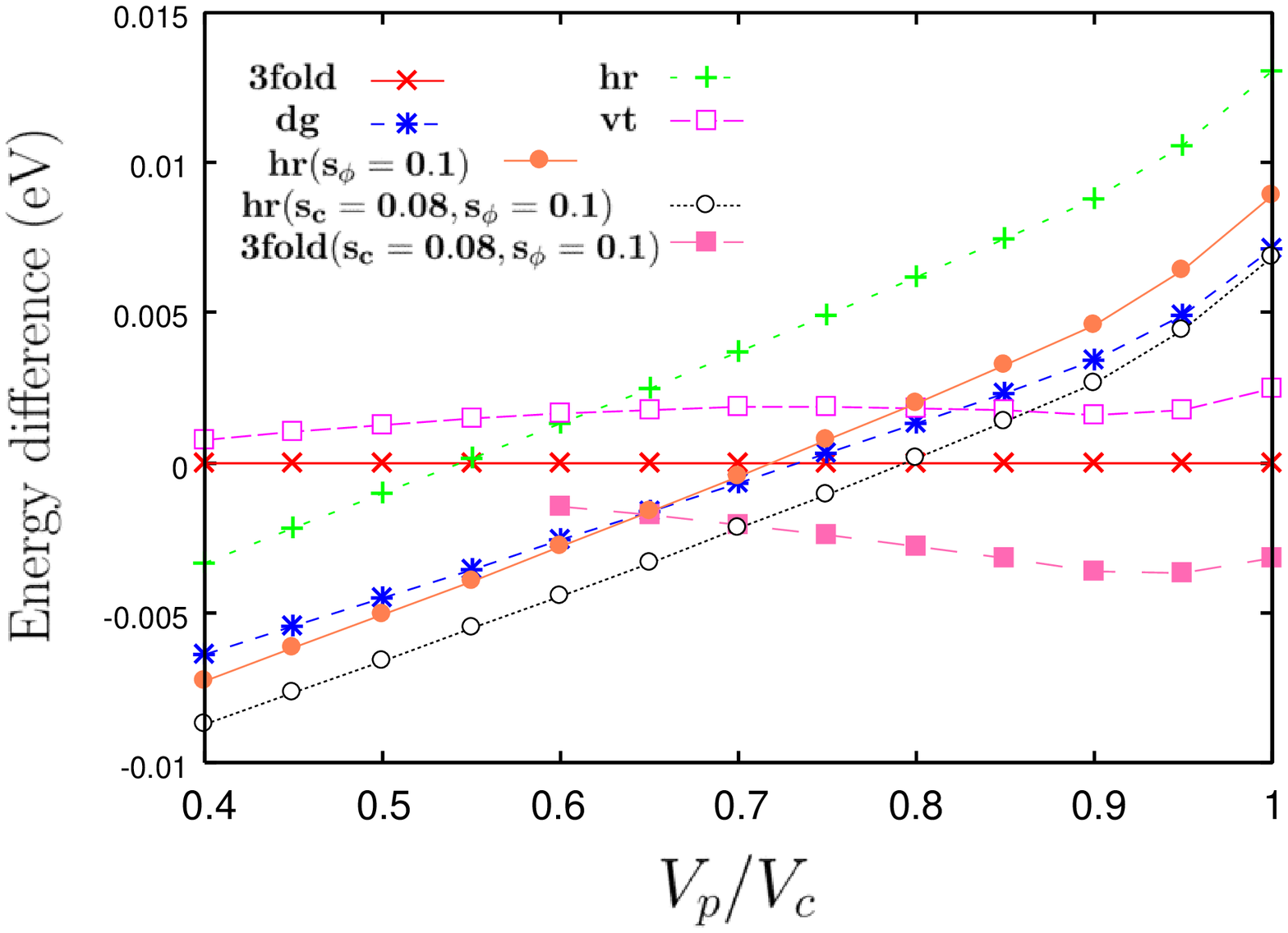}\\
\vspace{0.2cm}
\includegraphics[width=8.0cm,clip]{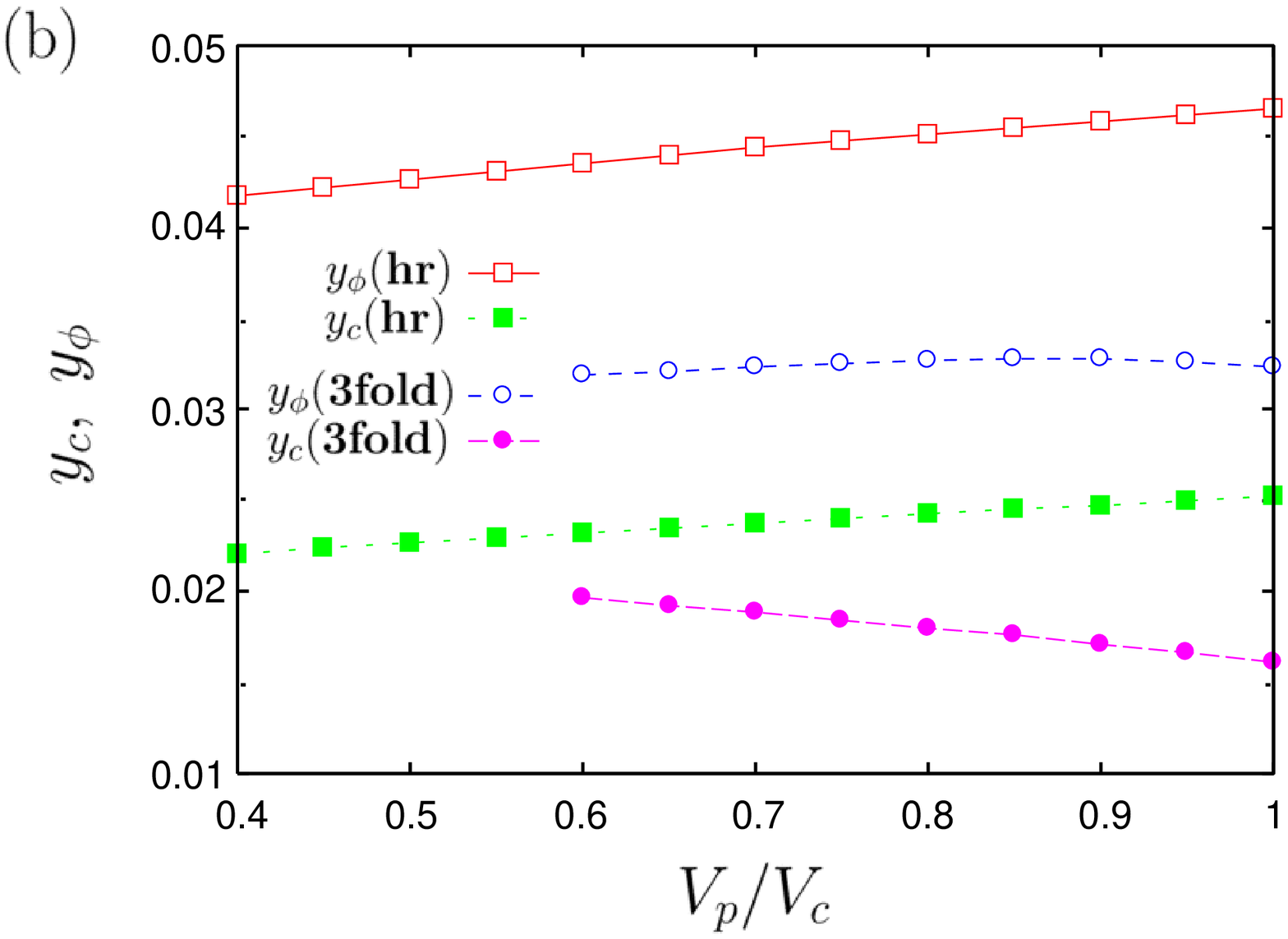}
\end{center}
\caption{(Color online) (a)Ground-state energies for various CO states
 as a function of
 the ratio $V_p/V_c$ for $U=0.7$ and $V_c/U=0.35$. The energy of 3-fold
 CO is chosen to be zero. hr, dg, and vt mean the horizontal, diagonal, and
 vertical COs, respectively. (b) Modulations in the transfer integrals
 for the horizontal and 3-fold COs.}
\end{figure}

Figure 7(a) shows that the electron-lattice couplings stabilize the
horizontal CO considerably. Here, we concentrate on the effects of
$s_{\rm c}$ and $s_{\phi}$ separately since the results in the case
where the three electron-lattice
couplings are simultaneously present have been shown
previously\cite{Tanaka}. The diagonal and vertical COs are not affected
by $s_{\rm c}$ and $s_{\phi}$, then the horizontal CO becomes more stable
than these states. The values of the electron-lattice couplings are
chosen to be $s_c=0.08$ and $s_{\phi}=0.1$.
The resultant lattice displacements are shown in Fig. 7(b),
which are comparable to the experimentally observed values.
The horizontal CO has hole-rich sites on the $t_{p4}$ chains,
which is consistent with the experimental findings. The 3-fold CO also
has energy
gain from the lattice modulations due to $s_{c}$ and $s_{\phi}$. To
obtain the mean-field solution for the 3-fold CO that coexists with
lattice modulations,
we have assumed the unit cell of $2\times6$ sites in the $a$-$c$ plane. 
In this state, a weak horizontal charge modulation is caused by the
lattice distortion in the background of the 3-fold CO. The
distortions in transfer integrals are weak compared to those in the
horizontal CO. Note that the 3-fold CO is metallic even if the transfer
integrals are modulated while the horizontal CO is insulating. In short,
the horizontal CO with lattice distortion becomes stable for
$V_p/V_c<0.7$, while the 3-fold CO is favorable for $0.7<V_p/V_c\lesssim
1$.

Here we briefly mention the effect of $s_{a}$. Since the $y_{a}$
distortion reduces the horizontal CO, $y_{c}$ and $y_{\phi}$ tend to be
suppressed when $y_{a}$ is large\cite{Tanaka}. The energy of the
horizontal CO is only slightly lowered by this $y_{a}$ distortion. The
relative stability of this CO compared to the others in the range of
$V_p/V_c<0.7$ is unchanged as long as we choose $s_{a}$ to obtain the
experimentally observed value of $y_a$.

At finite temperatures, the stability of COs is investigated by
calculating the free energy within the Hartree-Fock approximation. The
used values of $U$, $V_c/U$, and the electron-lattice couplings are the
same as those for $T=0$. The temperature dependences of the free
energies for various CO
patterns are shown in Figs. 8(a) and 8(b) in the cases of $V_p/V_c=1.0$
and $V_p/V_c=0.69$, respectively. For $V_p/V_c=1.0$, the 3-fold CO is 
the most stable state in a wide temperature range. On the other hand,
the horizontal CO with lattice modulations becomes more stable when
$V_p/V_c$ is small. In particular, a first-order metal-insulator
transition between the 3-fold CO and the horizontal CO occurs for
smaller values of $V_p/V_c$ than unity, as shown in Fig. 8(b) where the
transition is located at $T_c\sim 0.025$.
\begin{figure}
\begin{center}
\includegraphics[width=8.0cm,clip]{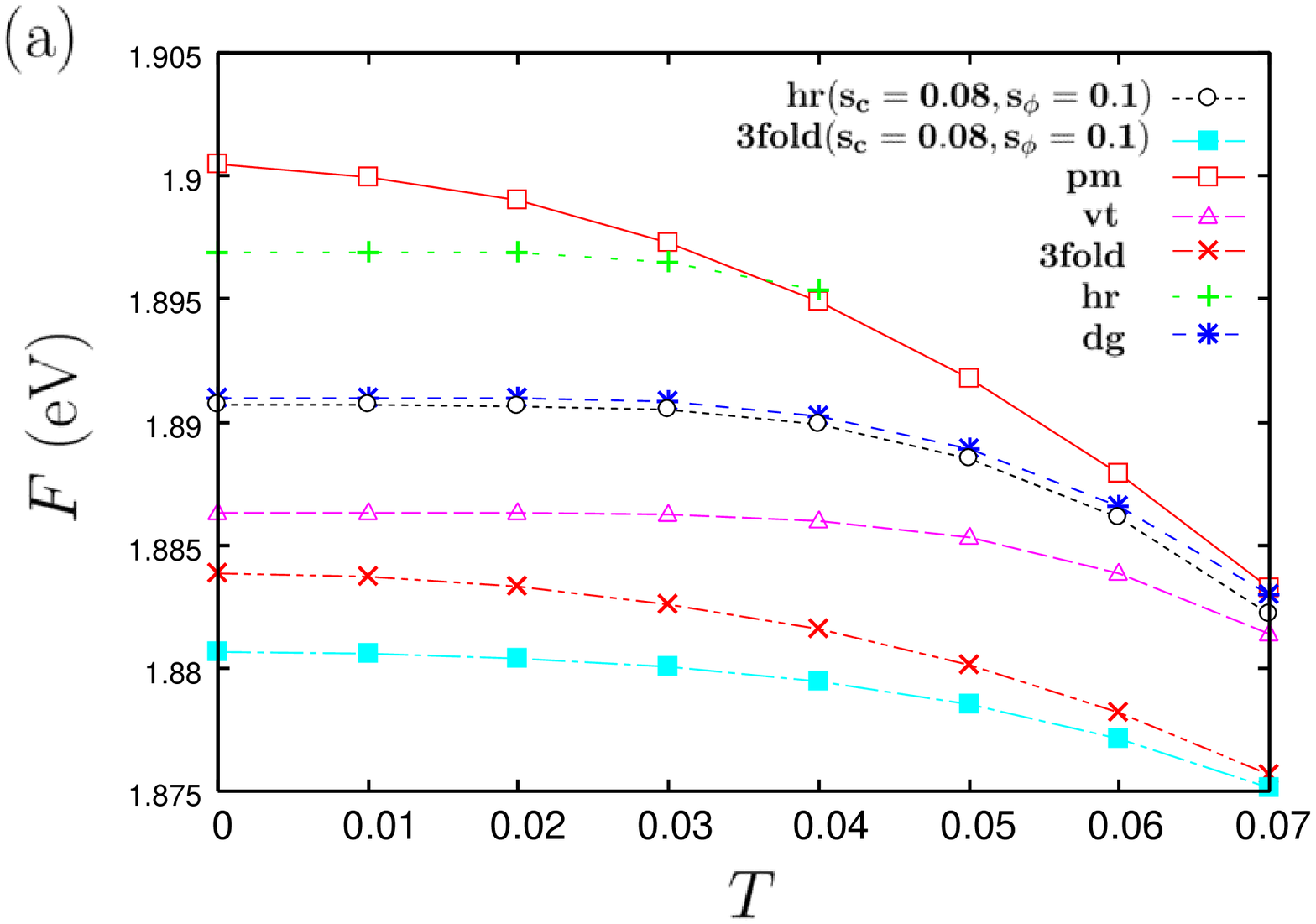}\\
\vspace{0.35cm}
\includegraphics[width=8.0cm,clip]{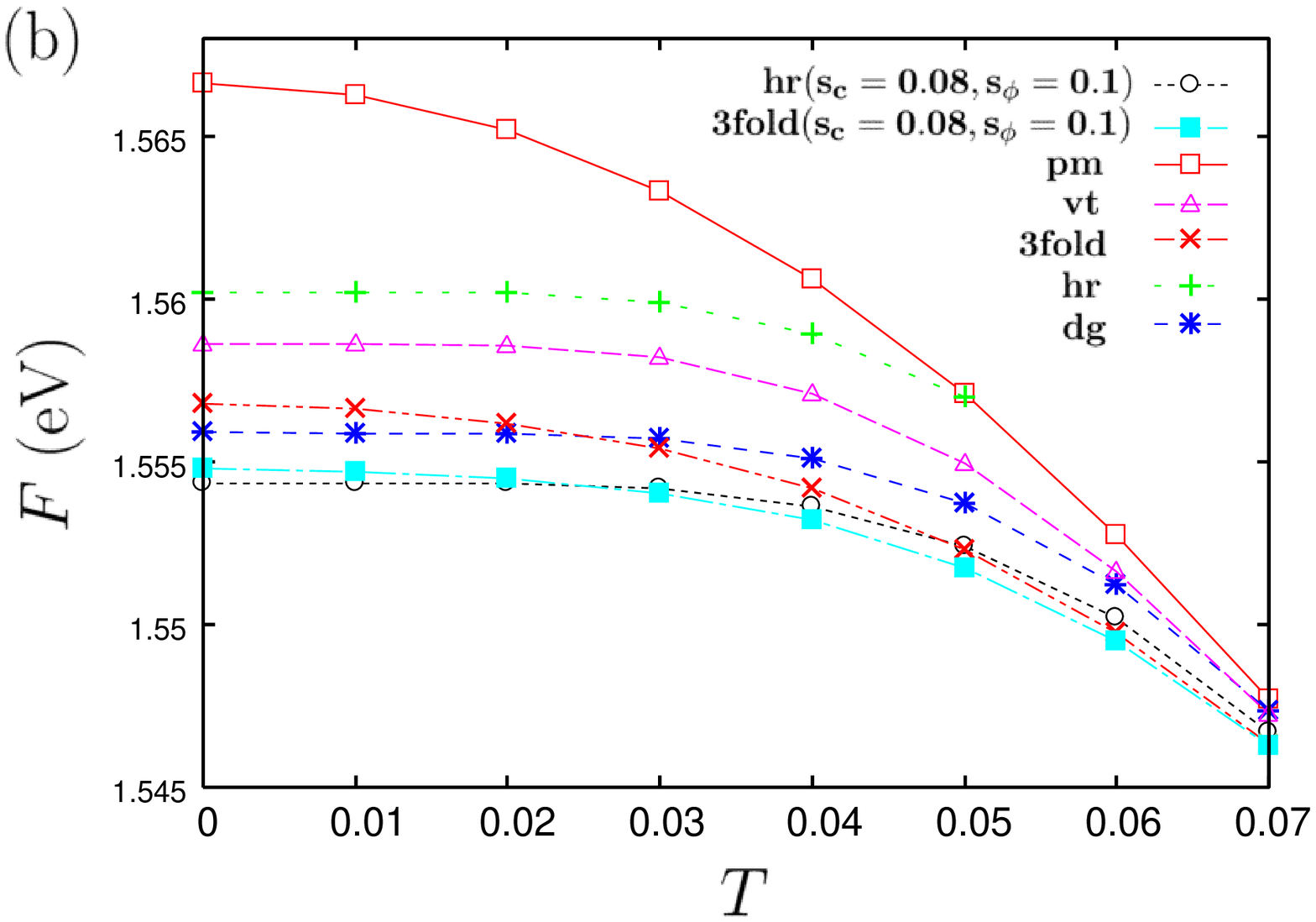}
\end{center}
\caption{(Color online) Free energies for various CO patterns for
 (a)$V_p/V_c=1.0$ and
 (b)$V_p/V_c=0.69$ in the case of $U=0.7$ and $V_c/U=0.35$. pm means the
 paramagnetic metallic state with  uniform charge density. A
 first-order metal-insulator transition from the 3-fold CO to the
 horizontal CO occurs at $T_c\sim 0.025$.}
\end{figure}

The first-order metal-insulator transition at a finite
temperature can be related to the experimental results of
$\theta$-(ET)$_2$RbZn(SCN)$_4$
although the wave vector of the charge modulation in the 3-fold CO state 
is different from that of the experiments in the
metallic phase. It has recently been proposed that 
the experimental observation can be reproduced by considering Coulomb
interactions extended to a longer range than the
nearest-neighbor sites\cite{Kuroki}. 
As for the spin degrees of
freedom, both the 3-fold and the horizontal COs in our Hartree-Fock
calculation have spin orders which have not been observed in the
experiments. The effect of quantum fluctuations is necessary in
discussing the behavior of the spin degrees of freedom.

The estimation of the intersite Coulomb interactions $V_p$
and $V_c$ indicates that these values are comparable\cite{Mori3},
$V_p/V_c\sim 1$, where the 3-fold CO is the most stable in our
calculation.
A variational Monte Carlo study\cite{Watanabe3} 
for the extended Hubbard model without electron-lattice couplings
also shows that the 3-fold CO is stable for 
$V_p/V_c\sim 1$.
According to the recent exact-diagonalization study\cite{Miyashita}, the
horizontal CO with lattice distortions becomes more stable even at
$V_p/V_c\sim 1$ if we take account of quantum fluctuations that are
neglected in the Hartree-Fock approximation. 
It is also indicated that the Holstein-type
electron-lattice coupling stabilizes the horizontal CO by an
exact-diagonalization study\cite{Clay}.

We find that the 3-fold state with coexisting weak horizontal charge
modulation is stable at the nearly isotropic region $V_p/V_c\sim 1$.
This can be related to the X-ray diffraction results on
$\theta$-(ET)$_2$CsZn(SCN)$_4$\cite{Watanabe4,Nogami}, which show two
types of COs coexisting as short-range fluctuations. 
Although the present Hartree-Fock calculation gives an artificial
long-range CO, quantum fluctuations are expected to destroy the
long-range order and result in a state where different types of COs
coexist as short-range fluctuations.

\subsection{$\alpha$-(ET)$_2$I$_3$}

As in the previous subsection, we first discuss the 
effects of two electron-lattice couplings $s_{\rm b1}$ and $s_{\rm b2}$
on the horizontal CO. In the CO state, sites A and B are hole-rich,
while sites C and D are
hole-poor. The spin configuration is antiferromagnetic along the stripe
and ferromagnetic between the stripes. We
choose $U=0.7$, $V_c/U=0.4$, and $V_p/V_c=0.8$
here. Figures. 9 and 10 show the lattice displacements and order
parameters for the horizontal CO in the presence of $s_{b1}$ and
$s_{b2}$, respectively. Although the charge distribution does not
change largely, the displacements 
increase with increasing the electron-lattice coupling in both cases,
which shows that these couplings stabilize the horizontal CO. This can
be seen from the energy gain due to the distortions. As shown in
Fig. 11, the energy of the horizontal CO is lowered in the presence of
$s_{b1}$ and $s_{b2}$.
\begin{figure}
\begin{center}
\includegraphics[width=8.0cm]{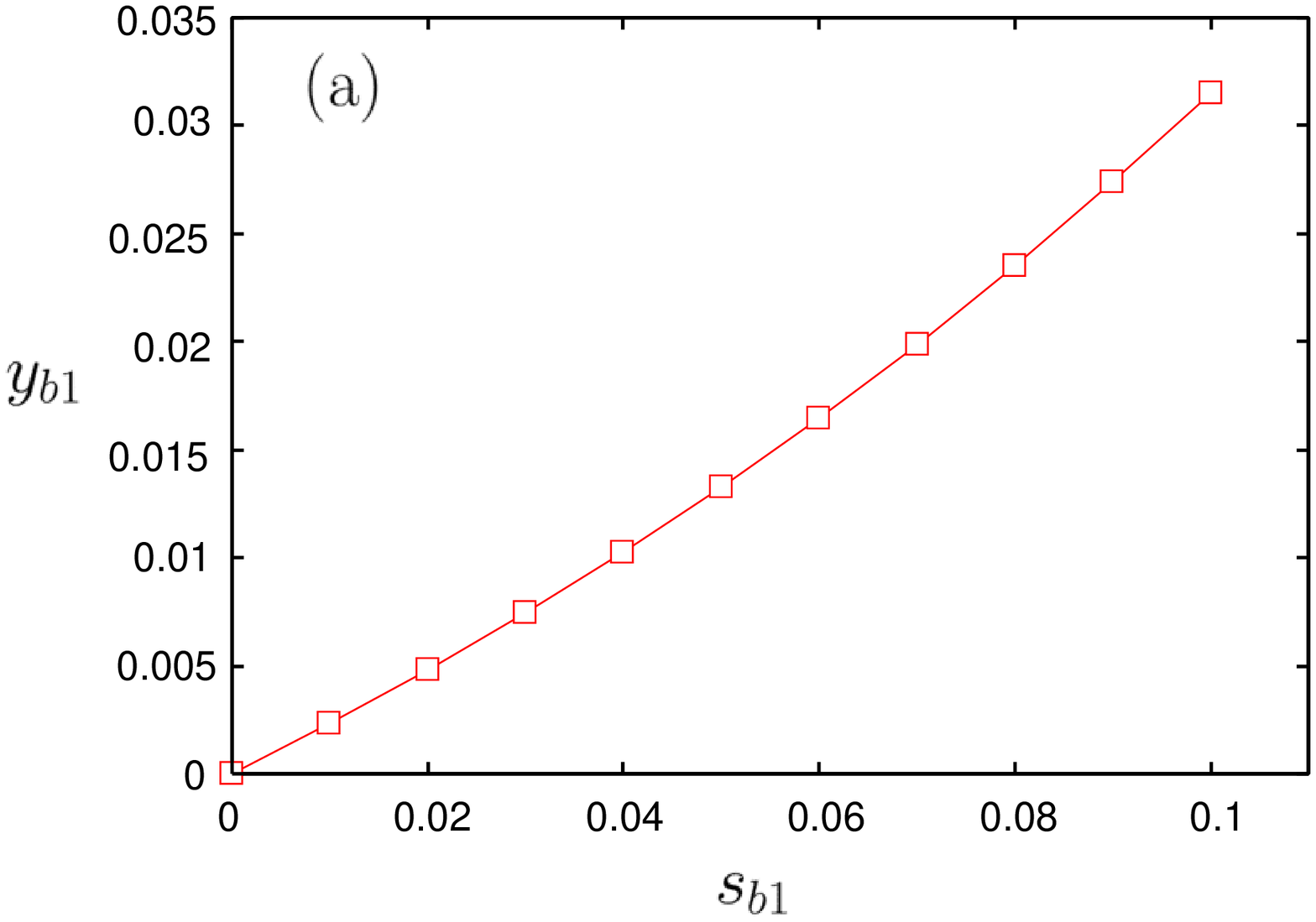}\\
\vspace{0.3cm}
\includegraphics[width=8.0cm]{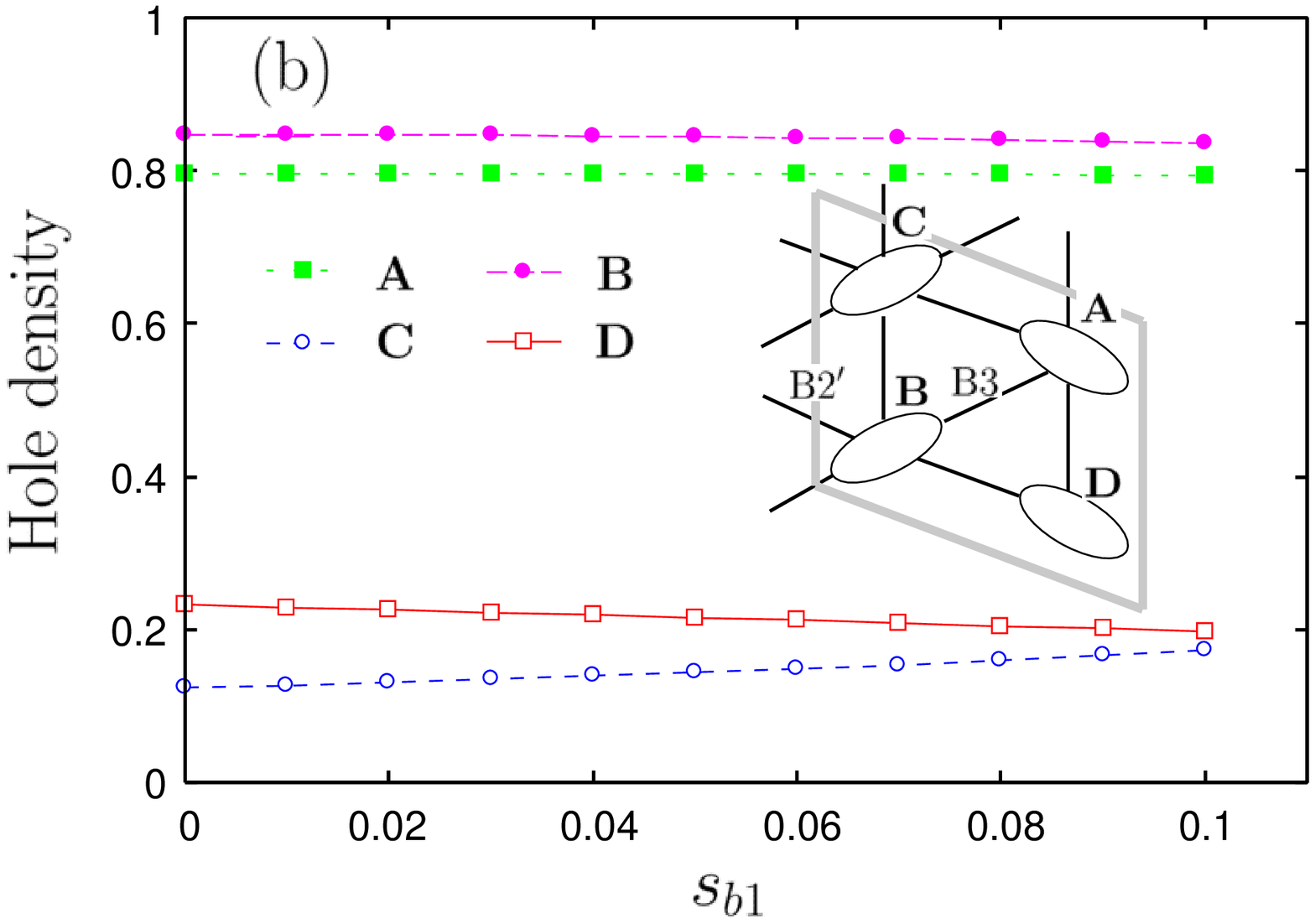}
\end{center}
\caption{(Color online) (a)Modulation $y_{b1}$ and (b)hole densities as
 a function of $s_{b1}$ for $U=0.7$, $V_c/U=0.4$ and $V_p/V_c=0.8$.}
\end{figure}
\begin{figure}
\begin{center}
\includegraphics[width=8.0cm]{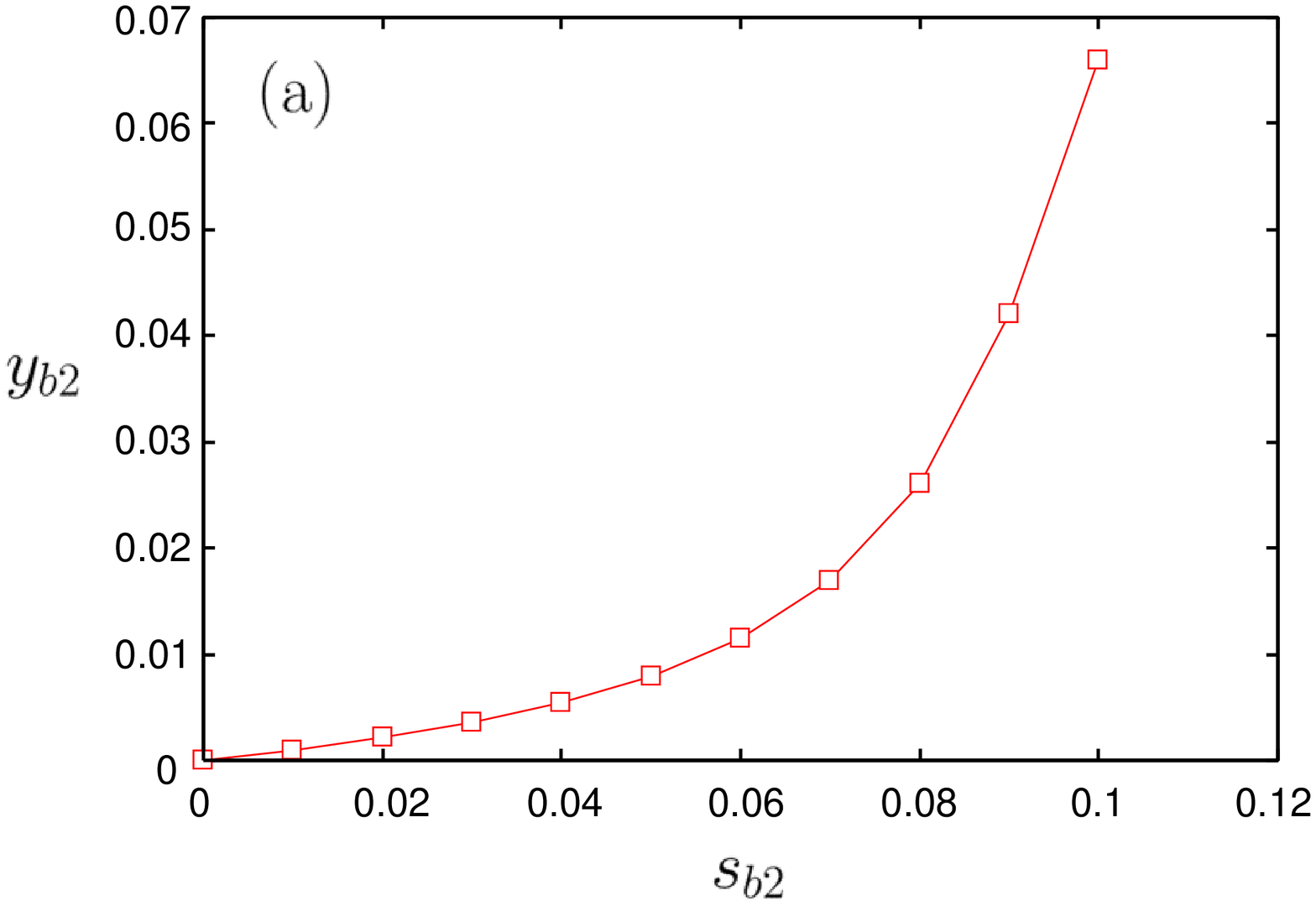}\\
\vspace{0.3cm}
\includegraphics[width=8.0cm]{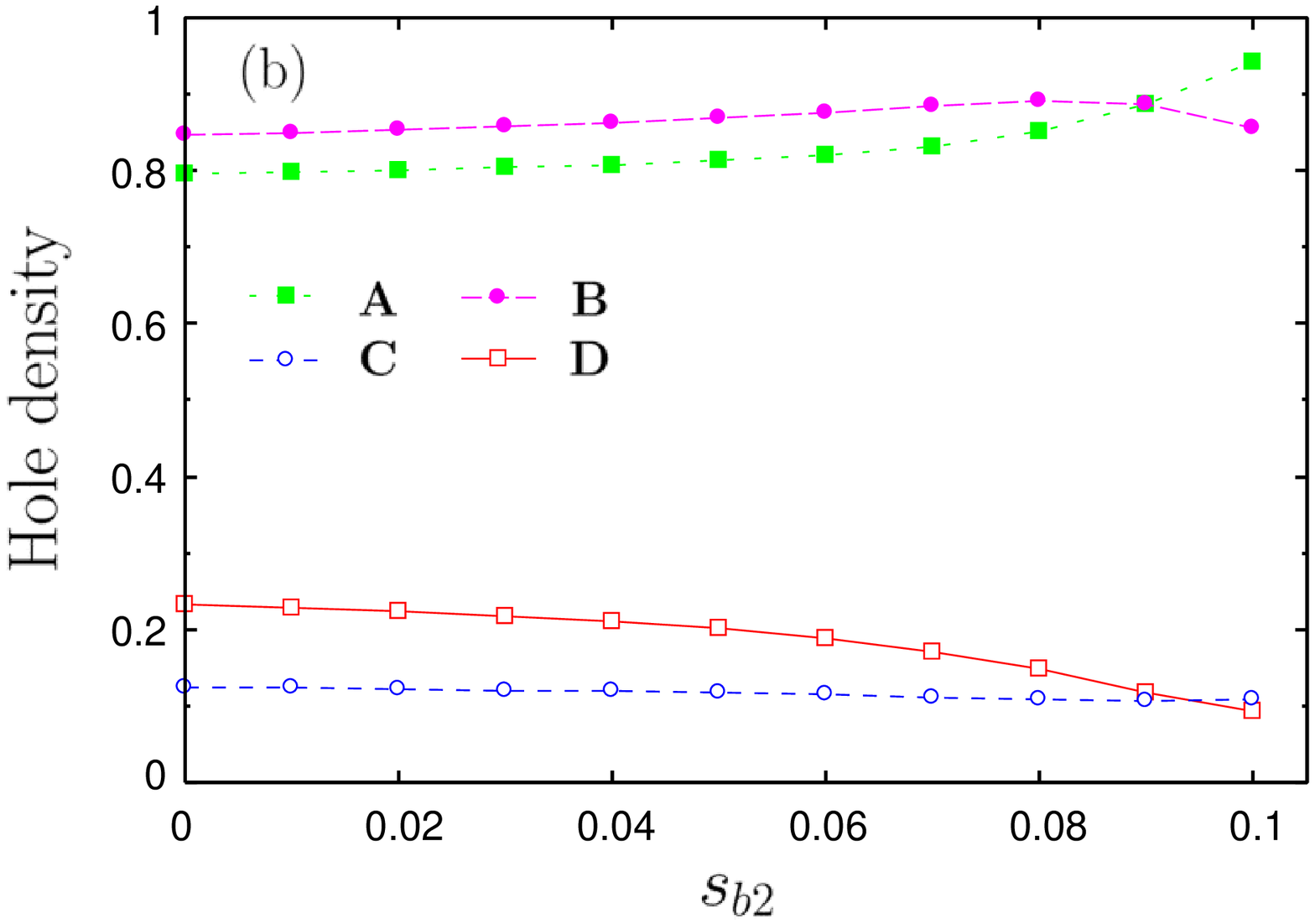}
\end{center}
\caption{(Color online) (a)Modulation $y_{b2}$ and (b)hole densities as
 a function of $s_{b2}$ for $U=0.7$, $V_c/U=0.4$ and $V_p/V_c=0.8$.}
\end{figure}
\begin{figure}
\begin{center}
\includegraphics[width=8.0cm]{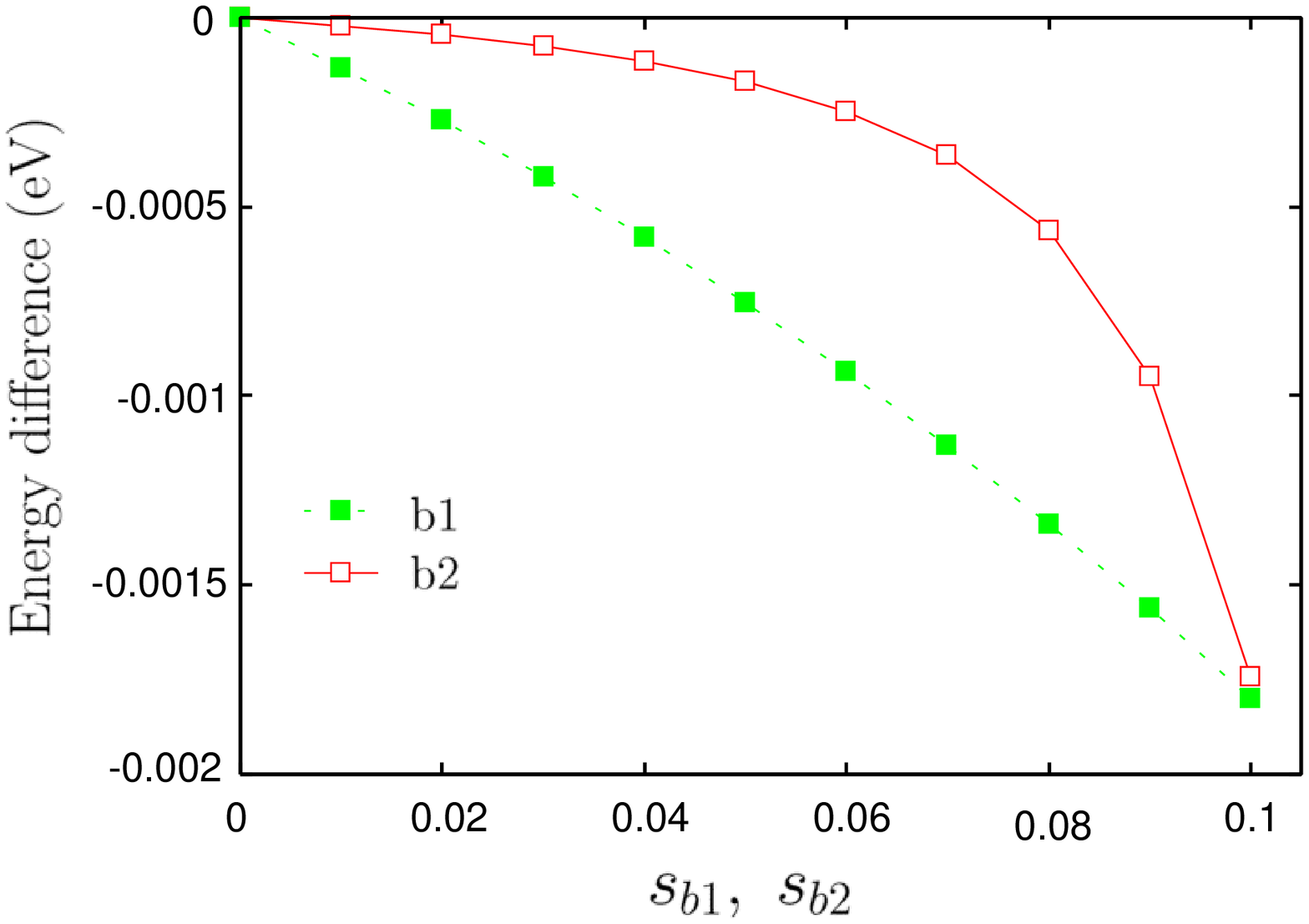}
\end{center}
\caption{(Color online) Energy gain due to each electron-lattice
 coupling in the horizontal CO for $U=0.7$, $V_c/U=0.4$, and
 $V_p/V_c=0.8$.}
\end{figure}

\begin{figure}[h]
\begin{center}
\includegraphics[width=8.0cm]{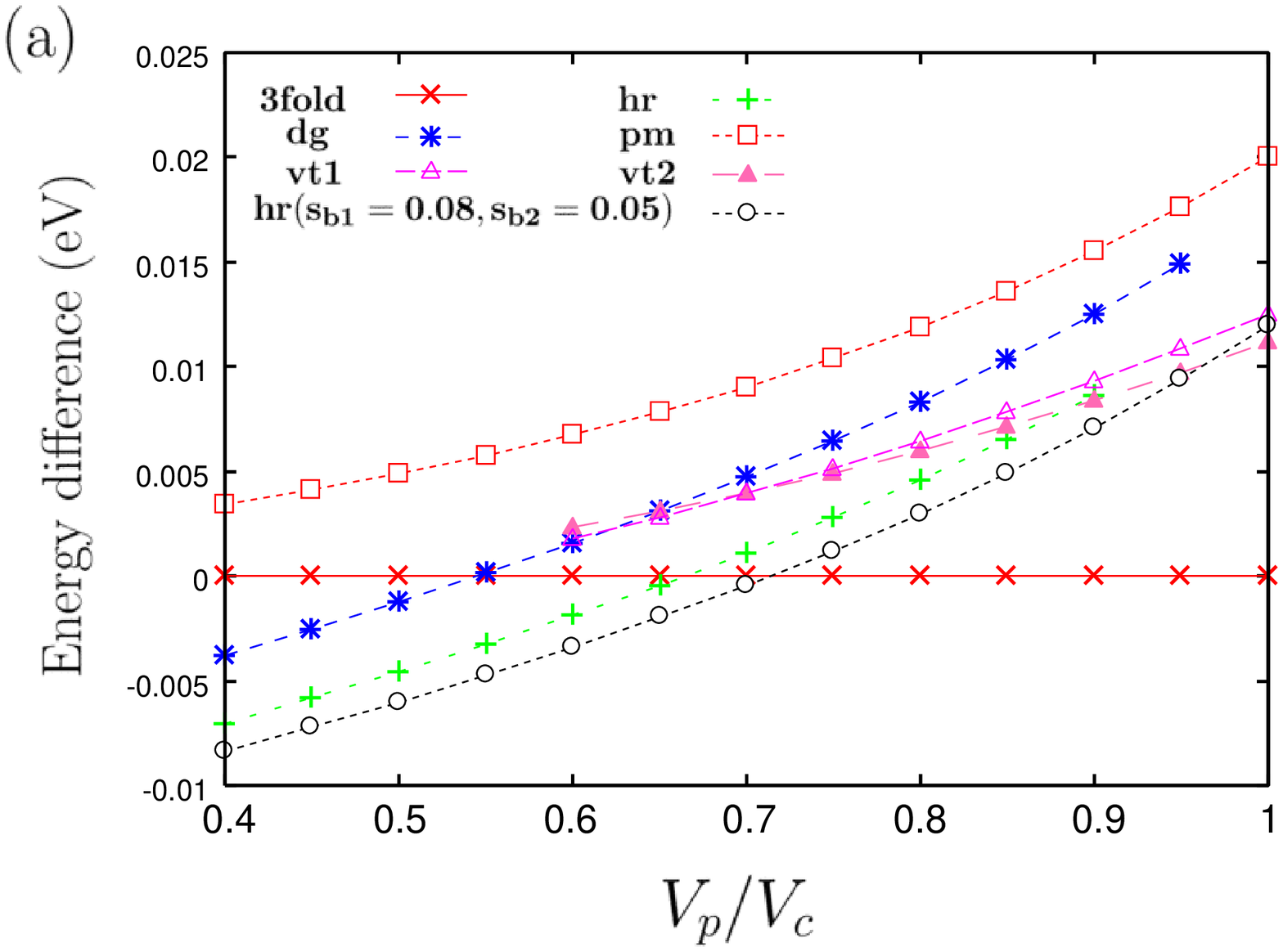}\\
\vspace{0.35cm}
\includegraphics[width=8.0cm]{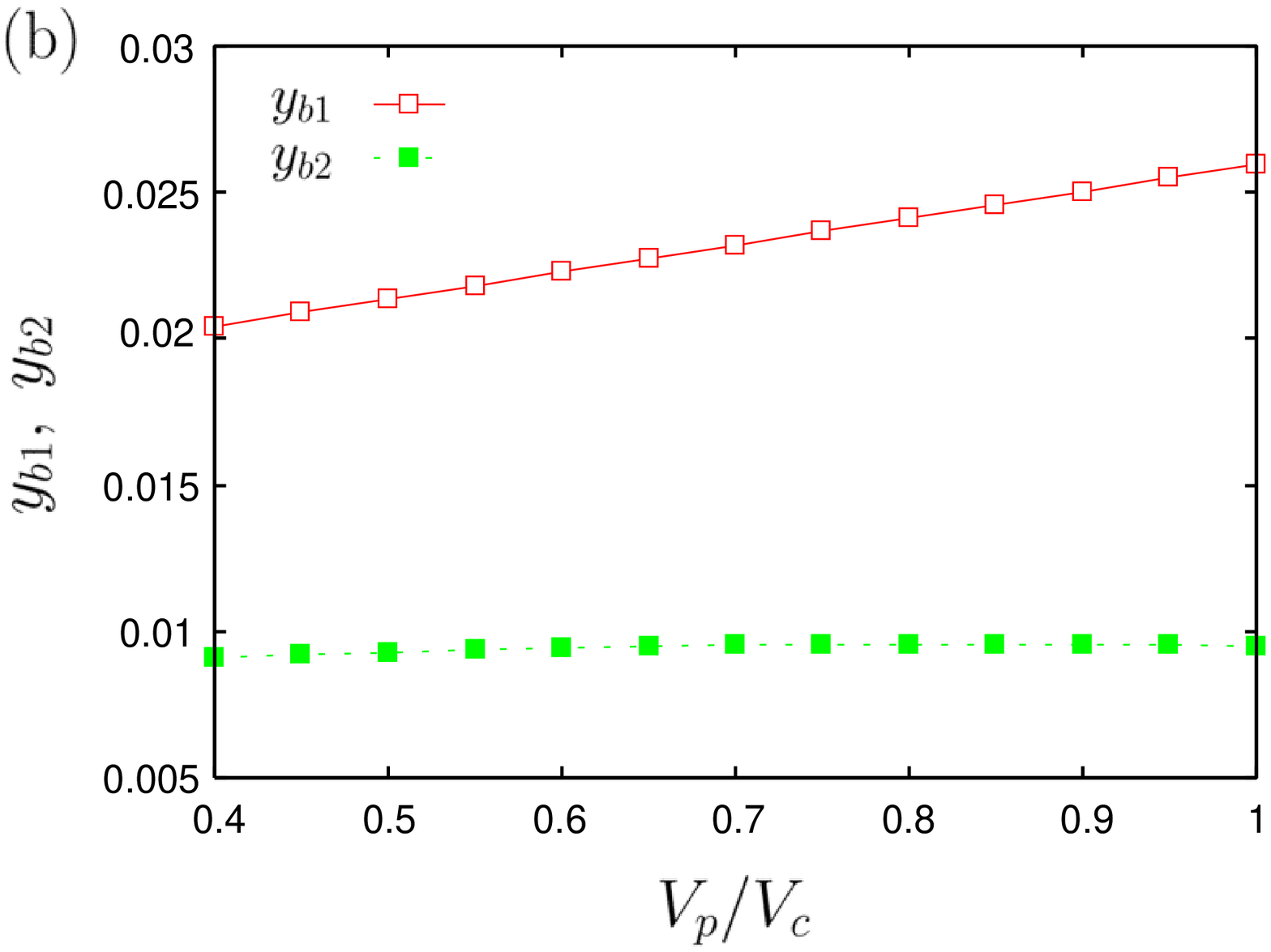}
\end{center}
\caption{(Color online) (a)Ground-state energies for various CO states
 as a function of the ratio $V_p/V_c$ for $U=0.7$ and $V_c/U=0.4$. 
(b)Modulations in the transfer integrals for the horizontal CO.}
\end{figure}
\begin{figure}[!h]
\begin{center}
\includegraphics[width=8.0cm]{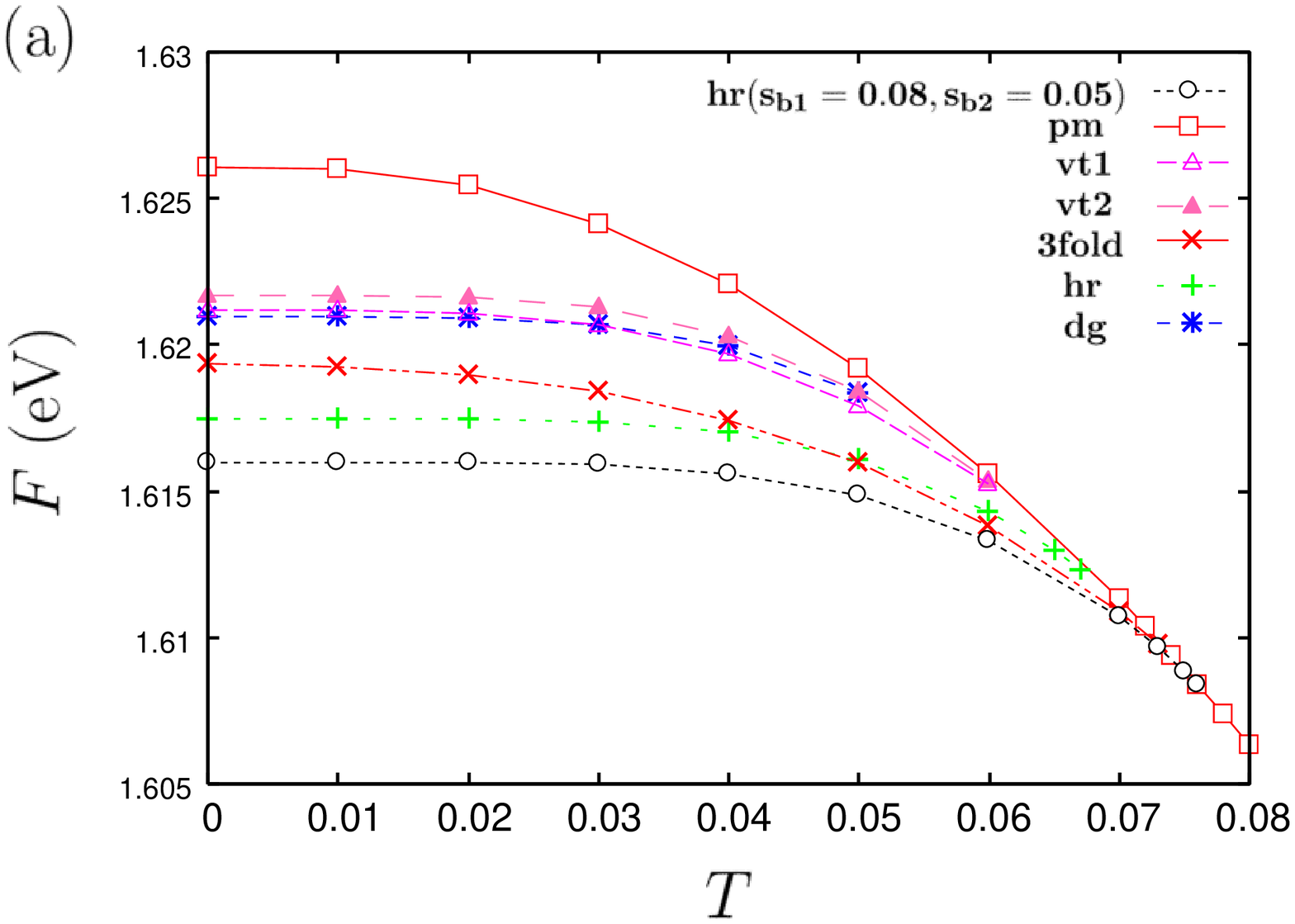}\\
\vspace{0.35cm}
\includegraphics[width=8.0cm]{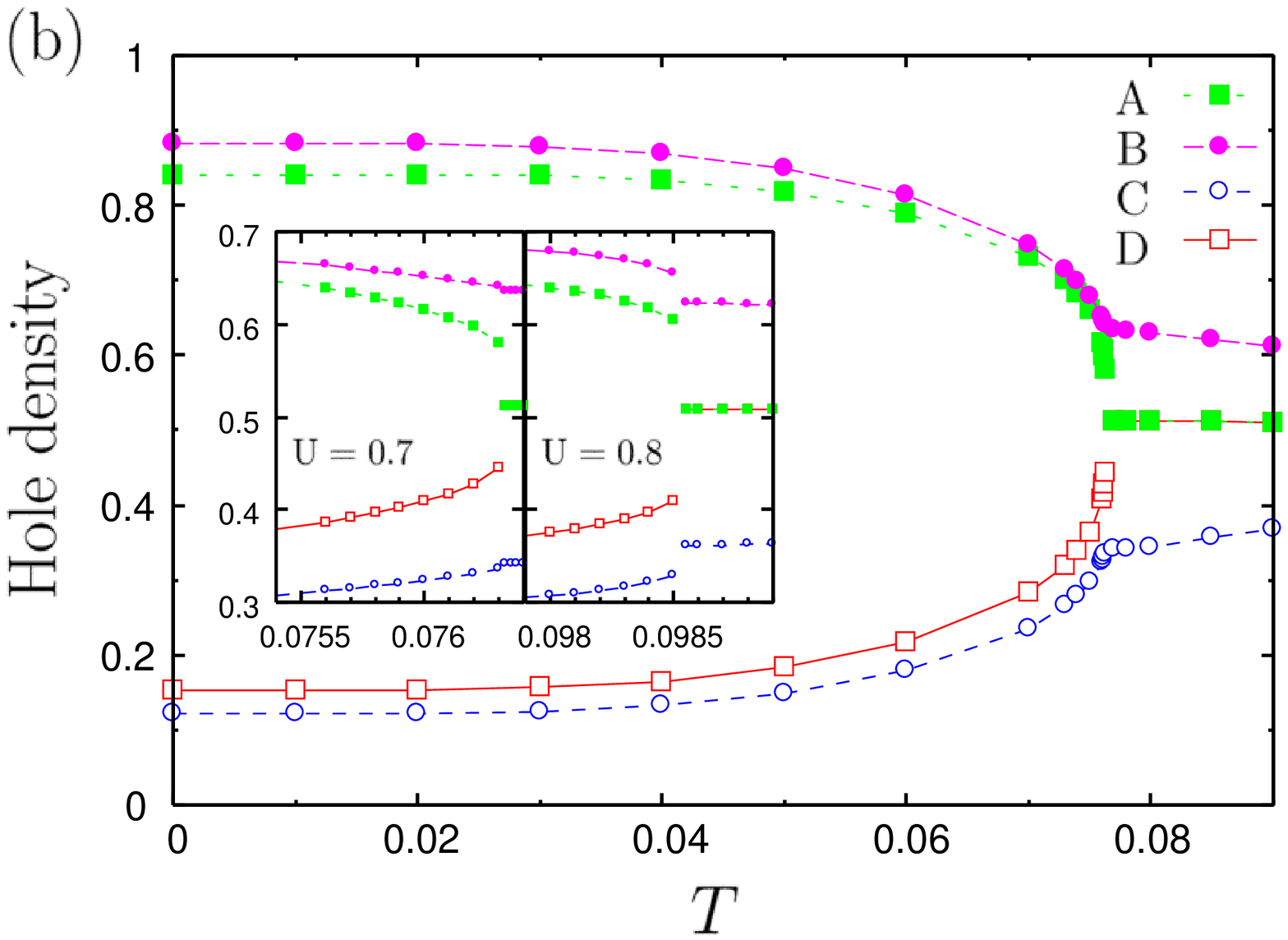}\\
\vspace{0.35cm}
\includegraphics[width=8.0cm]{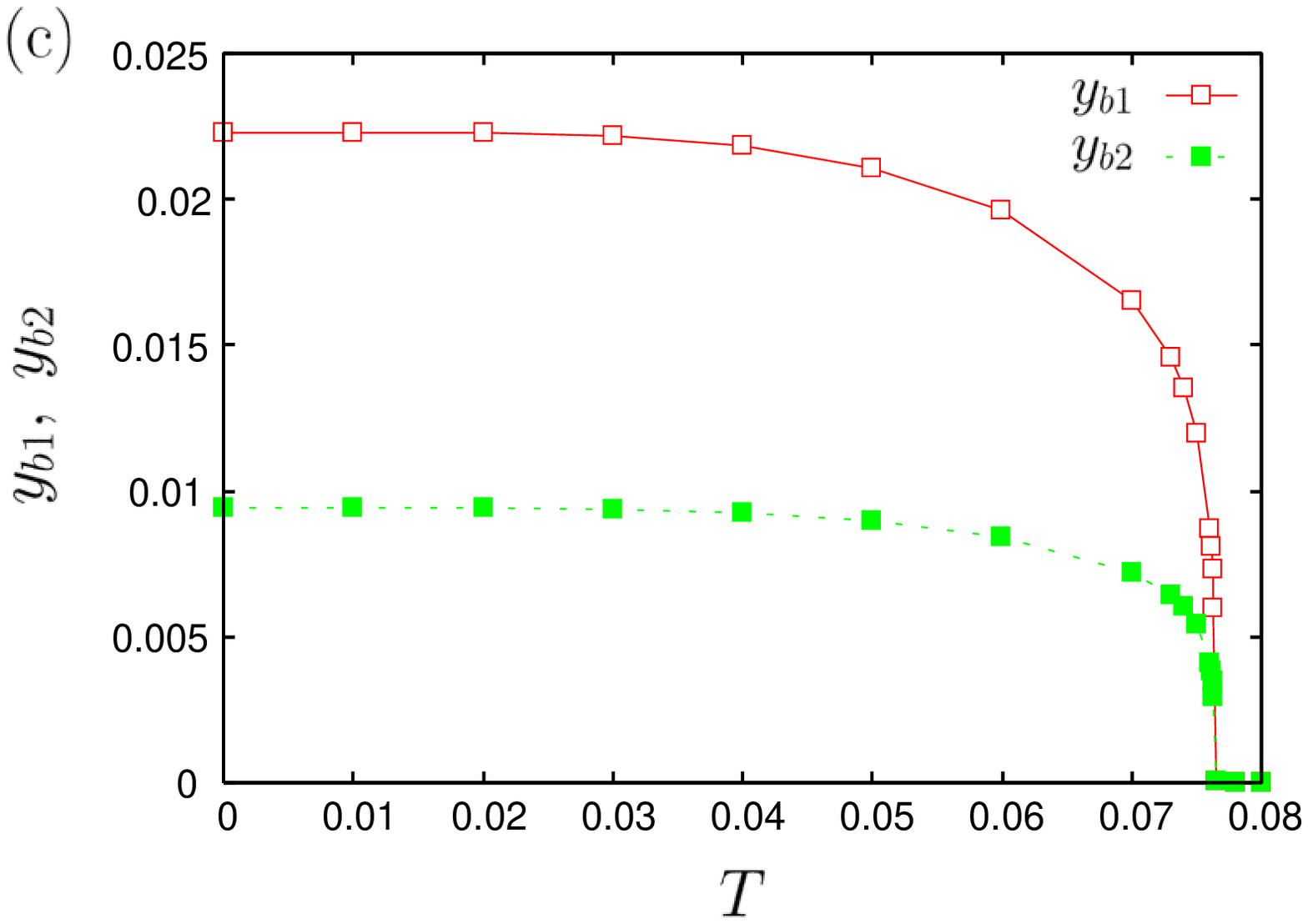}
\end{center}
\caption{(Color online) Temperature dependence of (a)free energies for
 various CO patterns, (b)charge density at each site, and (c)modulations
 $y_{b1}$ and $y_{b2}$ in the case of $U=0.7$, $V_c/U=0.4$, and
 $V_p/V_c=0.6$. The inset in (b) shows the behaviors of charge densities
 near the transition temperature for $U=0.7$ (left) and $U=0.8$ (right)
 with fixed values of $V_c/U=0.4$ and $V_p/V_c=0.6$.}
\end{figure}

The ground-state energies of various CO patterns per site are compared in
Fig. 12(a). We set $U=0.7$, $V_c/U=0.4$ here. The electron-lattice
couplings are chosen to be $s_{b1}=0.08$ and $s_{b2}=0.05$. The
3-fold CO has a ferrimagnetic spin configuration in the 2$\times$6 unit
cell as in the case of the $\theta$-type salt. For the vertical CO, the
energies of two 
solutions that are nearly degenerate are shown. vt1 means the vertical
CO with hole-rich sites A and D whereas vt2 denotes the one with
hole-rich sites B and C. For vt1, the spins on the stripes are
antiferromagnetic and those
between the stripes are ferromagnetic. vt2 has a ferromagnetic spin
configuration on the stripes and an antiferromagnetic one between the
stripes. It can be seen from Fig. 12(a) that the 3-fold CO is the most
stable near the isotropic repulsions, i.e., $V_p/V_c\sim 1$. When the
anisotropy in $V_p/V_c$ is large, i.e., $V_p/V_c<0.65$, on the other hand, the
horizontal CO is more stable
than the other CO states even in the absence of electron-lattice
couplings. This has been pointed out within the Hartree
approximation\cite{Seo}. This is in contrast to the $\theta$-type salt
where the horizontal CO does not become the
ground state without electron-lattice couplings. 
Therefore, the band structure which originates from the variety of
transfer integrals in the high temperature phase is considered to have
an important role to stabilize the CO in the $\alpha$-type salt. 
In the presence
of the electron-lattice couplings, the horizontal
CO is more stabilized and it becomes the ground state for
$V_p/V_c<0.7$. No other state is affected by these electron-lattice
couplings. In Fig. 12(b), we plotted the modulations in the transfer
integrals, which are consistent with the experimental values, i.e., 
$y_{b1}\sim 0.02$, and $y_{b2}\sim 0.01$\cite{Kakiuchi} in the
region where the horizontal CO becomes the ground state.
In the parameter range we have chosen, the paramagnetic metallic solution
has always a higher energy than the other CO patterns at $T=0$. In this
state, the charge distribution is not uniform due to the low symmetry of
the crystal structure\cite{Kobayashi,Kobayashi2}. Site B is hole-rich and
site C is hole-poor while the charge densities on sites A and D are the
same in the metallic phase. This state is the so-called zero-gap
semiconducting state where the energy band has two contact points at the
Fermi energy in the Brillouin zone\cite{Katayama,Kobayashi3}.

In above calculations, we do not show the modulations in transfer
integrals such as $t_{a1}$ and $t_{b4}$. In fact, we find that the
distortion in $t_{a1}$ slightly lowers the energy of the horizontal CO
although the distortion is smaller than those in $t_{b1}$ and $t_{b2}$.
For the change in $t_{b4}$, the experimentally observed
modulation does not become a self-consistent solution to the
Hartree-Fock calculation. This is due to the fact that the distortion
gains no energy since the increasing bond $t_{\rm B4^{\prime}}$
connects the hole-poor sites C and D. Therefore, we expect that the
changes in $t_{b1}$ and $t_{b2}$ have a major role in stabilizing the
horizontal CO and the obtained results are qualitatively unchanged by
other modulations.

Finally, we consider the case of finite temperatures. As discussed
above, the anisotropy in $V_p/V_c$ determines the stability of the CO
states as in the case of the $\theta$-type salt. For $V_p/V_c\sim 1.0$
the 3-fold state is stable, while the horizontal CO is stable when
$V_p/V_c$ is small. In the latter case, a phase transition occurs from
the paramagnetic metallic state to the horizontal CO. The
temperature dependence of the free energies of various CO patterns for
$V_p/V_c=0.6$
is shown in Fig. 13(a). We see that the transition takes place at 
$T_c\sim 0.076$. The temperature dependences of
the order parameters and lattice displacements are shown in Figs. 13(b)
and 13(c), respectively. In the paramagnetic metallic phase, the charge
densities on sites A and D are the same and the modulations are absent, 
while these sites are distinct from
each other in the horizontal CO phase accompanied by the modulations in
the transfer integrals. 
If Holstein-type electron-lattice couplings were present, the
charge disproportionation would make a difference among the site
energies at sites A, B and C, even in the metallic phase.
For $T>T_c$, the contact points
in the zero-gap
state deviate from the Fermi energy, accompanied with thermally blurred
small Fermi pockets. Below the transition temperature, the spins on the
stripes are weakly antiferromagnetic, and
the spin order grows towards $T=0$. The transition
is of first order although the discontinuity in the order parameters is
small. In the inset of Fig. 13(b), we show the behaviors of
hole-densities near $T_{c}$ for $U=0.7$ and $U=0.8$ by fixing the ratios
$V_c/U=0.4$ and $V_p/V_c=0.6$. 
The first-order transition is more
evident in the case of large Coulomb interactions.
Note that when we take much larger values of the electron-lattice couplings,
$T_c$ becomes higher and the transition from the paramagnetic
metal to the paramagnetic horizontal CO becomes continuous. 
We also find that, in the present case of $U=0.7$ the CO transition
becomes continuous without electron-lattice couplings, which indicates
that the lattice effect can alter the order of the transition.

A first-order transition from the charge disproportionated metallic state
to the horizontal CO accompanied by the lattice distortion is consistent 
with the experimental findings in $\alpha$-(ET)$_2$I$_3$.
As for the spin degrees of freedom, the antiferromagnetic spin order is
obtained in the calculated CO state whereas the spin gap is
experimentally observed below $T_c$ in the material\cite{Rothamael}. 
Although this result is due to the Hartree-Fock approximation which does
not take account of quantum fluctuations, the spin-singlet formation
below $T_c$ can be expected by the fact that the spins on the horizontal
stripes along $t_{{\rm B2}^{\prime}}$ and $t_{\rm B3}$ bonds form
alternating Heisenberg chains as discussed by Seo\cite{Seo}.
Experimentally, a CO with long
periodicity such as the 3-fold CO is not observed in this compound
although the present calculation shows that it is stable for
$V_p/V_c\sim 1.0$. As in the case of the $\theta$-type salt, 
the effects of fluctuations 
may further stabilize the horizontal CO even in the region of
$V_p/V_c\sim 1.0$.

\section{Summary and Conclusions}
In the present paper, we study the role of the lattice degrees of
freedom to the formation of CO in the quasi-two-dimensional organic
conductors $\theta$-(ET)$_2$RbZn(SCN)$_4$ and $\alpha$-(ET)$_2$I$_3$. 
By taking account of Peierls-type electron-lattice couplings that cause
the experimentally observed lattice modulation in each
compound, we
investigate the relevant extended Hubbard model within the Hartree-Fock 
approximation. It is found that the electron-lattice couplings stabilize 
the horizontal CO for both compounds,
which is consistent with the experimental observations. For
$\theta$-(ET)$_2$RbZn(SCN)$_4$, the effect of the lattice distortion is
crucial to realize the horizontal CO, since it does not
become the ground state without electron-lattice
couplings. 
The combined contributions of electron and lattice degrees of freedom
to the CO transition are in fact suggested by the recent dielectric
permittivity measurement\cite{Nad}.
Our results indicate that the $s_{\phi}$ distortion, which
corresponds to the molecular rotation, is the most
important among the three couplings in order to obtain the horizontal CO.
This result is also qualitatively consistent with the
exact-diagonalization study\cite{Miyashita} for eq. (1) on small
clusters. At finite temperatures, a first-order metal-insulator
transition is obtained, which is related to the CO transition in
$\theta$-(ET)$_2$RbZn(SCN)$_4$. For $\alpha$-(ET)$_2$I$_3$, on the other
hand, the horizontal CO is already stable if we consider the full band structure 
and the anisotropy in the nearest neighbor Coulomb interactions,
although the electron-lattice couplings further lower the energy of this
CO. We have shown a finite-temperature first-order
transition from the paramagnetic metallic state to the horizontal CO
with modulations in the transfer integrals.
This result agrees with the experimental findings on
$\alpha$-(ET)$_2$I$_3$. Although the antiferromagnetic spin order is
artificially obtained in the Hatree-Fock calculation, the spin gap
behavior, which is observed in the material, is indeed expected since
the exchange couplings between the neighboring spins on the horizontal
stripe are alternating as pointed out previously\cite{Seo}.

The contrastive role of the lattice degrees of freedom in each compound
would appear in the different natures of the CO phase
transitions. $\theta$-(ET)$_2$RbZn(SCN)$_4$ in the metallic phase has a
simple band structure with higher symmetry than that of
$\alpha$-(ET)$_2$I$_3$, and various COs compete with each other. When
the 3-fold CO fluctuations are dominant at high temperatures as
suggested by the present calculation, 
it is natural to expect that the transition to the horizontal CO
accompanies a large structural distortion because it requires the
considerable replacement of the charge patterns with different unit
cells. This results in the first-order transition with large
discontinuity as observed experimentally. On the
other hand, for $\alpha$-(ET)$_2$I$_3$, the metallic state has already
the charge disproportionation due to 
the complexity of transfer integrals.
Even in this 
metallic phase, site B is hole-rich while site C is hole-poor, which is 
common to the charge distribution in the horizontal CO. Because of this,
the CO can be realized by merely breaking the equivalence of charge
densities in sites A and D within the high-temperature unit cell. This
can lead to the first-order transition with small discontinuity relative
to that in the $\theta$-(ET)$_2$RbZn(SCN)$_4$. 

\section*{Acknowledgment}
The authors would like to thank H. Seo and S. Miyashita for helpful
discussions. This work was supported by Grants-in-Aid for Scientific
Research on Priority Area ``Molecular Conductors'' (No. 15073224), for
Scientific Research (C) (No. 19540381), for Creative Scientific Research
(No. 15GS0216), and the Next Generation Super Computing Project
(Nanoscience Program) from the Ministry of Education, Culture, Sports,
Science and Technology, Japan.


\begin{thebibliography}{99}
\bibitem{Ishiguro}
T.\ Ishiguro,\ K.\ Yamaji\ and G.\ Saito:\ {\it Organic
	Superconductors}\ (Springer-Verlag,\ Berlin,\ 1998)\ 2nd ed.
\bibitem{Seo_Rev}
H. Seo, C. Hotta and H. Fukuyama: Chem. Rev. {\bf 104} (2004) 5005.
\bibitem{Miyagawa}
K. Miyagawa, A. Kawamoto and K. Kanoda: Phys. Rev. B {\bf 62} (2000)
	7679.
\bibitem{Chiba}
R. Chiba, H. M. Yamamoto, T. Nakamura and T. Takahashi:
	J. Phys. Chem. Solids. {\bf 62} (2001) 389.
\bibitem{Takano1}
Y. Takano, H. M. Yamamoto, K. Hiraki, T. Nakamura and T. Takahashi:
	J. Phys. Chem. Solids {\bf 62} (2001) 393.
\bibitem{Takano2}
Y. Takano, K. Hiraki, H. M. Yamamoto, T. Nakamura and T. Takahashi:
	Synth. Met. {\bf 120} (2001) 1081.
\bibitem{Mori1}
H. Mori, S. Tanaka and T. Mori: Phys. Rev. B {\bf 57} (1998) 12023.
\bibitem{Tajima}
H. Tajima, S. Kyoden, H. Mori and S. Tanaka: Phys. Rev. B {\bf 62}
	(2000) 9378.
\bibitem{Wang}
N. L. Wang, H. Mori, S. Tanaka, J. Dong, and B. P. Clayman: J. Phys.:
	Condens. Matter {\bf 13} (2001) 5463.
\bibitem{Yamamoto}
K. Yamamoto, K. Yakushi, K. Miyagawa, K. Kanoda and A. Kawamoto:
	Phys. Rev. B {\bf 65} (2002) 085110.
\bibitem{Watanabe1}
M. Watanabe, Y. Noda, Y. Nogami and H. Mori: J. Phys. Soc. Jpn. {\bf 73}
	(2004) 116.
\bibitem{Watanabe2}
M. Watanabe, Y. Noda, Y. Nogami and H. Mori: J. Phys. Soc. Jpn. {\bf 74}
	(2005) 2011.
\bibitem{Chiba2}
R. Chiba, K. Hiraki, T. Takahashi, H. M. Yamamoto, and T. Nakamura:
	Phys. Rev. Lett. {\bf 93} (2004) 216405.
\bibitem{Watanabe4}
M. Watanabe, Y. Nogami, K. Oshima, H. Mori and S. Tanaka:
	J. Phys. Soc. Jpn. {\bf 68} (1999) 2654.
\bibitem{Nogami}
Y. Nogami, J. -P. Pouget, M. Watanabe, K. Oshima, H. Mori, S. Tanaka
and T. Mori: Synth. Met. {\bf 103} (1999) 1911.
\bibitem{Bender}
K. Bender, K. Dietz, H. Endres, H. W. Helberg, I. Hennig,
	H. J. Keller,H. W. Schafer and D. Schweitzer:
	Mol. Cryst. Liq. Cryst. {\bf 107} (1984) 45.
\bibitem{Rothamael}
B. Rothamael, L. Forro, J. R. Cooper, J. S. Schilling, M. Weger,
	P. Bele, H. Brunner, D. Schweitzer and H. J. Keller:
	Phys. Rev. B {\bf 34} (1986) 704.
\bibitem{Kino1}
H. Kino and H. Fukuyama: J. Phys. Soc. Jpn. {\bf 64} (1995) 1877.
\bibitem{Kino2}
H. Kino and H. Fukuyama: J. Phys. Soc. Jpn. {\bf 65} (1996) 2158.
\bibitem{Woj}
R. Wojciechowski, K. Yamamoto, K. Yakushi, M. Inokuchi, and A. Kawamoto:
	Phys. Rev. B {\bf 67} (2003) 224105.
\bibitem{Moroto}
S. Moroto, K.-I. Hiraki, Y. Takano, Y. Kubo, T. Takahashi,
	H. M. Yamamoto, and T. Nakamura: J. Phys. IV France {\bf 114}
	(2004) 399.
\bibitem{Kakiuchi}
T. Kakiuchi, Y. Wakabayashi, H. Sawa, T. Takahashi and T. Nakamura:
	J. Phys. Soc. Jpn. {\bf 76} (2007) 113702.
\bibitem{Emge}
T. J. Emge, P. C. W. Leung, M. A. Beno, H. H. Wang and J. M. Williams:
	Mol. Cryst. Liq. Cryst. {\bf 138} (1986) 393.
\bibitem{Endres}
H. Endres, J. Keller, R. Swietlik, D. Schweitzer, K. Angemund and
	C. Kruger: Z. Naturforsch. A {\bf 41} (1986) 1391.
\bibitem{Nogami2}
Y. Nogami, S. Kagoshima, T. Sugano and G. Saito: Synth. Met. {\bf 16}
	(1986) 367.
\bibitem{Nishio}
Y. Nishio: unpublished.
\bibitem{Fortune}
N. A. Fortune, K. Murata, M. Ishibashi, M. Tokumoto, N. Kinoshita and 
H. Anzai: Solid State Commun. {\bf 79} (1991) 265.
\bibitem{Iwai}
S. Iwai, K. Yamamoto, A. Kashiwazaki, F. Hiramatsu, H. Nakaya,
	Y. Kawakami, K. Yakushi, H. Okamoto, H. Mori and Y. Nishio:
	Phys. Rev. Lett {\bf 98} (2007) 097402.
\bibitem{Seo}
H. Seo: J. Phys. Soc. Jpn. {\bf 69} (2000) 805. 
\bibitem{Mckenzie}
R. H. McKenzie, J. Merino, J. B. Marston, and O. P. Sushkov:
	Phys. Rev. B {\bf 64} (2001) 085109.
\bibitem{Clay}
R. T. Clay, S. Mazumdar and D. K. Campbell: J. Phys. Soc. Jpn. {\bf 71}
(2002) 1816.
\bibitem{Mori2}
T. Mori: J. Phys. Soc. Jpn. {\bf 72} (2003) 1469.
\bibitem{Merino}
J. Merino, H. Seo and M. Ogata: Phys. Rev. B {\bf 71} (2005) 125111.
\bibitem{Kaneko}
M. Kaneko and M. Ogata: J. Phys. Soc. Jpn. {\bf 75} (2006) 014710.
\bibitem{Watanabe3}
H. Watanabe and M. Ogata: J. Phys. Soc. Jpn. {\bf 75} (2006) 063702.
\bibitem{Kuroki}
K. Kuroki: J. Phys. Soc. Jpn. {\bf 75} (2006) 114716.
\bibitem{Hotta1}
C. Hotta, N. Furukawa, A. Nakagawa and K. Kubo: J. Phys. Soc. Jpn. 
	{\bf 75} (2006) 123704.
\bibitem{Hotta2}
C. Hotta and N. Furukawa: Phys. Rev. B {\bf 74} (2006) 193107.
\bibitem{Udagawa}
M. Udagawa and Y. Motome: Phys. Rev. Lett. {\bf 98} (2007) 206405.
\bibitem{Tanaka}
Y. Tanaka and K. Yonemitsu: J. Phys. Soc. Jpn. {\bf 76} (2007) 053708.
\bibitem{Miyashita}
S. Miyashita and K. Yonemitsu: Phys. Rev. B {\bf 75} (2007) 245112.
\bibitem{Watanabe5}
M. Watanabe: private communication.
\bibitem{Mori4}
T. Mori, A. Kobayashi, Y. Sasaki, H. Kobayashi, G. Saito, and
	H. Inokuchi: Chem. Lett. {\bf 13} (1984) 957.
\bibitem{Mori3}
T. Mori: Bull. Chem. Soc. Jpn. {\bf 73} (2000) 2243.
\bibitem{Kobayashi}
A. Kobayashi, S. Katayama, K. Noguchi and Y. Suzumura:
	J. Phys. Soc. Jpn. {\bf 73} (2004) 3135.
\bibitem{Kobayashi2}
A. Kobayashi, S. Katayama, and Y. Suzumura:
	J. Phys. Soc. Jpn. {\bf 74} (2005) 2897.
\bibitem{Katayama}
S. Katayama, A. Kobayashi, and Y. Suzumura:
	J. Phys. Soc. Jpn. {\bf 75} (2006) 054705.
\bibitem{Kobayashi3}
A. Kobayashi, S. Katayama, Y. Suzumura, and H. Fukuyama:
	J. Phys. Soc. Jpn. {\bf 76} (2007) 034711.
\bibitem{Nad}
F. Nad, P. Monceau, and H. M. Yamamoto: Phys. Rev. B {\bf 76} (2007)
	205101.
\end{thebibliography}
\end{document}